\documentclass[sigconf,nonacm]{acmart}

\AtBeginDocument{%
  }

\usepackage{multirow}   
\usepackage{array}       
\usepackage{xcolor}       
\usepackage{colortbl}  
\usepackage{graphicx} 

\usepackage{booktabs,tabularx,ragged2e}
\newcolumntype{Y}{>{\RaggedRight\arraybackslash}X}

\usepackage[most]{tcolorbox}

\usepackage{float}

\begin{document}

\title{Durably Reducing Belief in Women's Health Misinformation Through Culturally Adaptive AI Videos}

\author{Anku Rani}
\affiliation{%
  \institution{MIT Media Lab}
  \institution{Massachusetts Institute of Technology}
  \city{Cambridge}
  \state{MA}
  \country{USA}
}
\email{ankurani@mit.edu}

\author{Kokil Jaidka}
\affiliation{%
  \institution{Department of Communications and New Media}
  \institution{National University of Singapore}
  \city{Singapore}
  \country{Singapore}
}

\author{Shruti Sharma}
\affiliation{%
  \institution{Social Impact Group}
  \institution{Tata Power}
  \city{New Delhi}
  \country{India}
}

\author{Pragya Mahajan}
\affiliation{%
  \institution{Social Impact Group}
  \institution{Tata Power}
  \city{New Delhi}
  \country{India}
}

\author{Manisha Wadhwa}
\affiliation{%
  \institution{Social Impact Group}
  \institution{Tata Power}
  \city{New Delhi}
  \country{India}
}

\author{Andrew B. Lippman}
\affiliation{%
  \institution{MIT Media Lab}
  \institution{Massachusetts Institute of Technology}
  \city{Cambridge}
  \state{MA}
  \country{USA}
}

\author{Pattie Maes}
\affiliation{%
  \institution{MIT Media Lab}
  \institution{Massachusetts Institute of Technology}
  \city{Cambridge}
  \state{MA}
  \country{USA}
}

\author{Paul Pu Liang}
\affiliation{%
  \institution{MIT Media Lab}
  \institution{Massachusetts Institute of Technology}
  \city{Cambridge}
  \state{MA}
  \country{USA}
}

\renewcommand{\shortauthors}{Rani et al.}

\begin{abstract}

Health misinformation disproportionately harms women, yet interventions rarely address the community norms that sustain false beliefs. We test whether culturally adaptive AI-generated video in which the presenter looks like someone from her community reduces misinformation belief among low-literacy women in suburban India. In a field experiment (N=434), participants watched an AI-generated video featuring either an adaptive or neutral presenter. The culturally adaptive presenter reduced misinformation belief by 30\%, nearly twice the reduction produced by the neutral presenter compared to the non-intervention control condition. Post-experiment interviews suggest women recalled the neutral condition as a generic video but recognized the adaptive presenter. Gains persisted for three weeks. The adaptive advantage was largest for beliefs reinforced by community, such as blaming women for infertility, and negligible for medical knowledge gaps, such as understanding vaccines. These findings demonstrate the potential of culturally adaptive AI interventions to counter socially embedded health misinformation.

\end{abstract}

\begin{teaserfigure}
 \includegraphics[width=\textwidth]{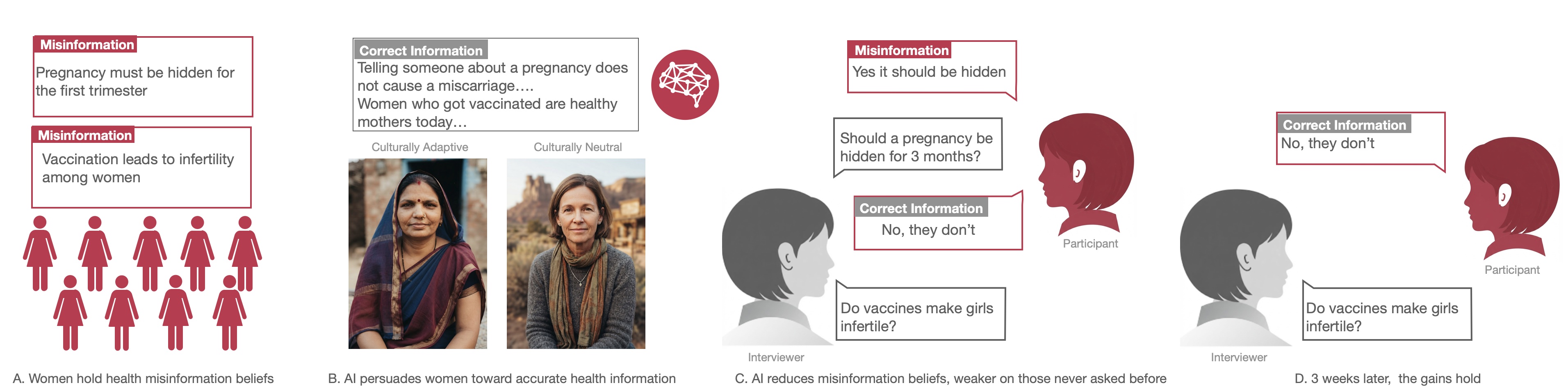}
 \caption{\textbf{Study overview.} Women answered questions on health misinformation (A), watched an AI video differing only in the presenter's cultural identity (B), and were retested immediately (C) and at two and three weeks (D). Cultural adaptation roughly doubled the immediate reduction in misinformation belief (30\% vs.\ control), and the advantage persisted over time.}
  \Description{A four-panel diagram read left to right. Panel A shows nine red female icons beneath two boxes labeled ``Misinformation'': ``Pregnancy must be hidden for the first trimester'' and ``Vaccination leads to infertility among women.'' Panel B shows a ``Correct Information'' box with an excerpt of the video script, an AI brain icon, and two presenter photos side by side: a culturally adaptive presenter, a South Asian woman in a sari, and a culturally neutral presenter, a woman in a grey sweater and scarf. Panel C shows a grey interviewer silhouette facing a red participant silhouette. The interviewer asks ``Should a pregnancy be hidden for 3 months?'' and ``Do vaccines make girls infertile?'' Speech bubbles show a misinformed answer, ``Yes it should be hidden,'' and a correct answer, ``No, they don't.'' Panel D shows the same pair three weeks later, with the participant correctly answering ``No, they don't'' to the vaccine question.}
  \label{fig:teaser}
\end{teaserfigure}

\maketitle

\section{Introduction}
Nearly six billion people, 74\% of the world, are now online \cite{itu2025facts}, and much of what reaches them about their health is wrong. A systematic review of 69 studies found health misinformation on every platform examined, most often about vaccines, where false posts made up as much as 43\% of the content and reached 87\% of users on some topics \cite{suarez2021prevalence}. Women bear a disproportionate share of this harm, yet efforts to counter health misinformation seldom target them. Misleading claims about contraception, fertility, maternal health, and vaccines are pervasive online \cite{john2025online}, which subsequently raise vaccine hesitancy and delay care \cite{donascimento2022infodemics}, and reduce uptake of essential services, including contraception and HPV vaccination \cite{purnat2025impacts}. Women are also the harder audience to reach with a written correction. Nearly two-thirds of the world's 739 million illiterate adults are women \cite{unesco2025literacy}; 71\% of women use the internet against 77\% of men, a gap of 280 million people that has not narrowed since 2019 \cite{itu2025facts}; and 810 million women in low- and middle-income countries do not use mobile internet at all, two-thirds of them in South Asia and sub-Saharan Africa \cite{gsma2026gendergap}. Research on mitigating misinformation, like fact-checking itself, follows the connected and the literate. Its field experiments have concentrated on political rather than health claims \cite{badrinathan2021educative, guess2020digital}, and fact-checkers describe their own reach as urban, text-based, and falling off sharply in resource-constrained settings \cite{seelam2024fact}.
The women most exposed to health misinformation are therefore the least represented in, and minimal beneficiaries of, existing corrective interventions.
This study was conducted in India, where in 2025 the country had 958 million active internet users, 57\% of them rural and 47\% women, and 588 million of them, three in five, watched short video formats that year, with rural viewers slightly outnumbering urban ones \cite{ptindtv2026internet}. The home-grown short-video platforms that serve this audience draw most of their users from small towns and villages, who watch for about half an hour a day, much of it in regional languages \cite{redseer2024india}. For women who reach the internet on a family handset, video is the medium: they make up 58\% of rural India's shared-device users \cite{ptindtv2026internet}, and in the community of low-literacy suburban Indian women that we study, 62\% watch YouTube daily. Yet among Indians aged 15--49, only 72\% of women are literate compared to 84\% of men, 33\% of women have ever used the internet versus 57\% of men, and 54\% of women have a phone they themselves use \cite{nfhs5}. Health misinformation is rampant in India: in an analysis of 419 items collected from an Indian fact-checking site, health was the most common theme of fake news, ahead of religion and politics, and nearly all of it had spread through four social platforms \cite{alzaman2021fakenews}. Community health workers are the intended last mile for these rural and suburban, low-literacy women, but they must persuade rather than merely inform, and they do so against resistance in the community and with limited training of their own \cite{10.1145/1753326.1753610}. 

Any correction must also deal with the sociocultural roots of health knowledge in these communities. In India, beliefs that vaccines cause infertility, that a pregnancy must be concealed for its first three months, that cloth is as safe as a sanitary napkin, or that childlessness is a curse to be resolved by a faith healer rather than a clinician persist across many low-income communities and
carry direct consequences for care-seeking \cite{jadhav2023treatment,chakrabarty2023assessing,josten2026barriers, khan2021my}. Such beliefs are sustained less by a shortage
of correct information than by traditional norms and by silence around women's health. Menstruation
has long remained a conversational taboo across India \cite{tuli2019sa}, and stigma suppresses the
very exchange through which these claims would ordinarily be challenged \cite{10.1145/3772318.3791318, chowdhury2025literaturerewadiscussingreproductive}. Misinformation about women's health therefore hardens into a norm, and the women who most need correction are the least able to seek it.

Video is the channel that does reach them. It is the largest single category of downstream internet traffic on both fixed and mobile networks, at 38\% of the total, and on-demand streaming, led by YouTube, accounts for 54\% of downstream volume \cite{sandvine2024phenomena}. Video requires neither literacy nor a device of one's own to consume, and a long line of work in Human Computer Interaction for Development (HCI4D) shows that speech and video, not text, carry health information to low-literate audiences (Section~\ref{sec:delivery}). We therefore posited that an intervention delivered in this medium, by a presenter the viewer recognizes as one of her own, offers a distinctive opportunity to correct health misinformation. Generated video makes such an intervention feasible at scale. Models such as Veo~3~\cite{GoogleDeepMind2025Veo3}, Sora~\cite{openai2024videoworldsimulators}, and
Wan~2.2~\cite{wan2025wanopenadvancedlargescale} produce photorealistic footage of a person
speaking, with synchronized audio, from a text prompt alone, at a cost low enough to scale to many variants of one message \cite{zhang2025generative, qian2025vc, chen2025code2video}. An intervention can therefore be delivered in a viewer's own language, dress, and setting without filming a presenter for every community. Whether that adaptation is worth its cost is unresolved: face-to-face culturally adapted health interventions show a moderate benefit \cite{griner2006culturally}, yet adapted internet- and mobile-based interventions may not repay the effort \cite{balci2022culturally}, and work on agent and avatar similarity finds that demographic matching reliably changes how people regard a messenger without reliably changing what they learn or believe \cite{baylor2004pedagogical, wang2026similar, lee2026health}. This raises the question we study:
\emph{Can a culturally adaptive AI video that looks and speaks like someone from a woman's own community change what she believes about her health, and does that change last and extend beyond the specific misinformation she is pre-asked before the intervention?}

\begin{figure*}
  \centering
  \includegraphics[width=\textwidth]{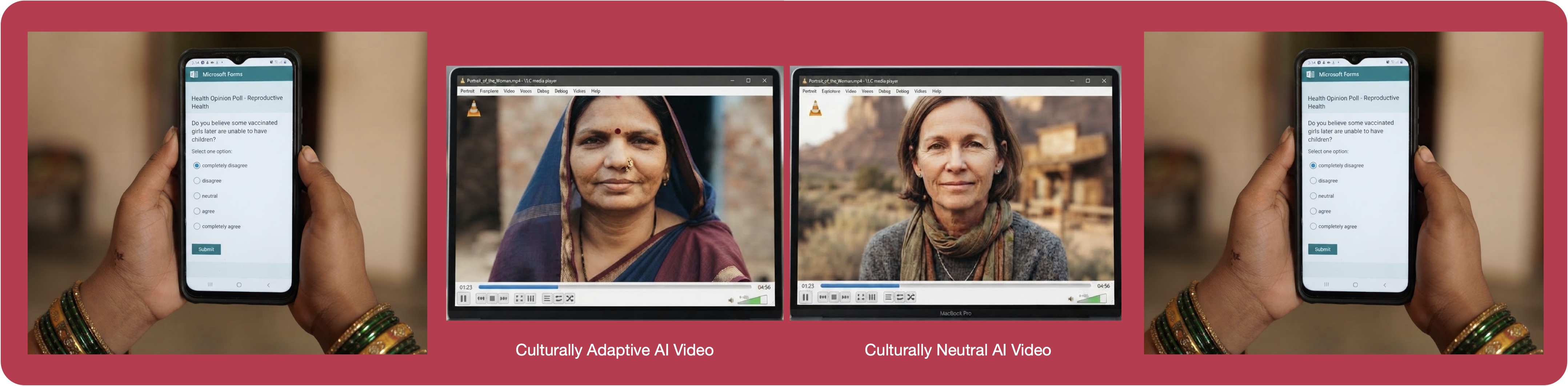}
  \caption{\textbf{Interaction flow.} Participants rate their belief in misinformation, watch a culturally adaptive or neutral AI video, and then rate their belief in misinformation again.}
  \label{fig:interaction}
  \Description{Three-panel sequence on a dark red background. Left: hands with bangles hold a smartphone showing a five-point survey question asking whether vaccinated girls later become unable to have children. Center: a laptop plays a video of a South Asian woman presenter in a sari. Right: the same survey question shown again on the smartphone.}
\end{figure*}

We propose culturally adaptive AI health videos: generated videos in which a presenter matched to the appearance of the viewer's community delivers persuasive, evidence-based health information, rigorously designed to combat health misinformation collected from the communities they address. Generating these videos is considerably harder. State-of-the-art generators render non-Western subjects and settings with uneven cultural faithfulness \cite{rani2026culturescoreevaluatingculturalfaithfulness}, and they produce only 4--8 second clips, so a twelve-minute video drifts in appearance across segments and cuts speech mid-clause. To keep the presenter looking the same throughout, we generate every clip from one reference photo. To keep her speech sounding natural, we split each script at natural pauses and size each piece to a normal Hindi speaking pace. We then join the 93 separately generated clips into one smooth video that plays on a laptop in the field. In the neutral condition, everything is kept constant apart from the appearance of the presenter, who looks like a woman from the West (Refer to Figure \ref{fig:interaction}).
We validate the result in a month-long randomized field study with low-literacy women recruited from women's literacy centers in Indian urban slums. Women's literacy centers are run by a local organization where women from low-literacy communities come to learn basic skills like counting or reading a bus number. We summarize our contributions as follows:

\begin{enumerate}

    \item \textbf{System design.} We present a pipeline that turns misinformation identified through formative field interviews with doctors about women's health into a persuasive, AI-generated video that guides women toward accurate health information. 
    We curate the misinformation list from fieldwork rather than published sources, since no such dataset exists, so the videos address what women in these communities actually believe.
    Figure \ref{fig:interaction} illustrates the flow of interaction.

    \item \textbf{Field study.} We conduct a month-long, three-condition randomized field study with $N = 434$ low-literacy women: a culturally adaptive AI presenter ($n = 188$), a culturally neutral AI presenter ($n = 162$), and a control ($n = 84$) group where women rate their belief in misinformation without any intervention. The two video conditions are identical in script, sequence, voice register, and runtime, and differ only in the presenter's appearance. Beliefs were measured by an interviewer-administered oral survey before and after viewing the video and again two and three weeks later. Each survey adds items the participant has never seen, so we can separate correction of beliefs she was asked about earlier from correction of beliefs she was never asked about.

    \item \textbf{Results.} On a belief score bounded at $\pm 1$, the culturally adaptive condition gained $+0.12$ belief score points (on a $-1$ to $+1$ scale, where $+1$ is full rejection of the misinformation) within the session against $+0.06$ for the neutral video condition. Adjusting for baseline belief score, the difference between conditions was significant ($p = .004$, $d = 0.31$), and both ended well above the no-video control ($d = 0.80$ and $d = 0.52$). The advantage over the neutral condition did not transfer: on claims never raised before viewing, the two were indistinguishable, and the culturally adaptive condition closed only 16\% of the gap to full rejection of misinformation, compared with 30\% on the claims it had been asked about beforehand.
    Both conditions retained their gains, and the adaptive advantage was undiminished three weeks later.
    
    \item \textbf{Open source code and dataset.} We will publicly release the generation pipeline code, both AI videos mitigating the misinformation, the women's health misinformation dataset derived from our formative interviews, and the de-identified survey responses, to enable replication and further study.

\end{enumerate}

\section{Related Work}

\subsection{Why women's health misinformation persists}
\label{sec:persist}
\label{sec:norms-vs-knowledge}

Claims that sanitary napkins cause infection or that vaccines cause infertility recur across resource-constrained settings \cite{10.1145/3772318.3791318, zacce2022impact}, and the fact-checking infrastructure that exists elsewhere is oriented toward literate, connected, urban users \cite{seelam2024fact}. Beliefs about women's bodies carry a further burden of silence: menstruation remains a conversational taboo across India \cite{tuli2019sa}, stigma suppresses the exchange through which such claims would ordinarily be challenged \cite{chowdhury2025literaturerewadiscussingreproductive}, and asking is itself socially costly when it reveals ignorance \cite{chandrasekhar2018signaling}. These social mechanisms compound the familiar cognitive ones, continuing to influence after retraction and truth by repetition \cite{doi:10.1177/1529100612451018, fazio2019repetition}.

The distinction that organizes our results is between two ways a false belief is held. An \emph{information deficit} persists because nobody has explained what a vaccine does, and any credible source closes it. A \emph{social norm}, in Bicchieri's sense, persists because a woman believes her reference group expects it \cite{bicchieri2017norms}, and it changes when perceived expectations change rather than when facts do: correcting Saudi men's underestimate of their peers' support for women working raised the share who enrolled their wives in a job-matching service \cite{bursztyn2020misperceived}. Mass media has shifted norms of this kind largely by changing who is seen holding a belief, from cable television and telenovelas to edutainment series and randomly assigned women leaders \cite{jensen2009power, laferrara2012soap, banerjee2019entertaining, beaman2009powerful}. Whether a generated presenter who merely looks like a member of the reference group can do the same is the question our design isolates.

\subsection{Correcting Health Misinformation in Low-Literacy Settings}
\label{sec:delivery}

Fact-checks and warning labels reduce belief in false claims modestly \cite{hartwig2024landscape, chuai2026community}, dialogue that engages a person's own reasons does so durably \cite{costello2024durably}, and inoculation videos establish video as a scalable carrier \cite{roozenbeek2022psychological}, with effects lasting about a month \cite{maertens2025psychological}. Nearly all of this evidence is online and high-literacy \cite{grover2024online}, and the field experiments that moved it to South Asia mostly disappointed: an hour of in-person media-literacy training in Bihar had no effect \cite{badrinathan2021educative}, and tip sheets and educational videos fared little better \cite{guess2020digital, ali2023countering, blair2024interventions}. What has worked is dose and source: a months-long classroom curriculum with rural adolescents held at four months \cite{amar2026countering}, and WhatsApp corrections from a trusted civil-society organization raised knowledge and changed behaviour \cite{bowles2020countering}.

HCI4D established speech and video as the channels through which health information reaches people for whom text is not one \cite{sherwani2007healthline, cuendet2013videokheti}, with human-mediated video the most durable model \cite{gandhi2007digital, abate2023accelerating, gurley125impacts, murthy2020effects}; in these settings ICTs persuade more than they inform \cite{10.1145/1753326.1753610}, and recent interventions target health workers rather than the communities whose beliefs resist them \cite{10.1145/3359271, 10.1145/3706598.3713680}. The rural-India immunization trials treat belief as a black box, moving vaccination records with reliable camps, incentives, and locally trusted ambassadors \cite{banerjee2010improving, banerjee2025selecting, banerjee2011poor}. Our study opens the box, at the cost of measuring stated belief rather than records (Section~\ref{sec:limitations}).

\subsection{Synthetic Messengers and Messenger Similarity}

Embodied agents have long served low-health-literacy users well \cite{10.1145/1518701.1518891, bickmore2016improving, 10.1145/3472306.3478350, jiang2024embodied}, and generative models now make photorealistic presenters cheap \cite{haider2025artificial, singh2025empowering}. Disclosure does not neutralize them: AI-generated video continued to shape belief even when viewers were warned it was synthetic \cite{clark2026continued}, which makes an openly generated health presenter viable, with narrative identification as the expected mechanism \cite{moyer2008toward}. Generators do, however, render South Asian subjects through an outsider's gaze \cite{qadri2023ai, rani2026culturescoreevaluatingculturalfaithfulness}, so community validation of any adapted video is a prerequisite.

Whether the messenger's identity changes what people believe is contested; work on agent and avatar similarity finds that demographic matching reliably changes how a messenger is regarded without reliably changing what is learned or believed \cite{baylor2004pedagogical, wang2026similar, lee2026health}, and adapted digital health promotion often does not repay its cost \cite{balci2022culturally}, though face-to-face adaptation shows moderate benefit \cite{griner2006culturally}. Health communication explains the split as \emph{surface-structure} tailoring, matching face, dress, and setting, which mainly buys attention, against \emph{deep-structure} tailoring, which engages the values sustaining a belief \cite{resnicow1999cultural, kreuter2003achieving}. Field experiments with human messengers point the other way: Black men assigned to Black doctors took up more preventive care \cite{alsan2019diversity}, lay messengers matched experts in raising vaccine demand \cite{alsan2024experimental}, and short videos sent to 25 million phones in West Bengal changed reported behavior \cite{banerjee2020messages, alsan2021comparison, torres2021effect, banerjee2019using, alatas2019celebrities}. Those studies observe behavior, with human messengers, mostly on information deficits; the avatar studies observe ratings, in laboratories, with literate participants. We test which side low-literacy health belief falls on: a pure surface-structure change to a generated presenter, script held constant, in a low-literacy population, against a no-video control, with belief measured before viewing, immediately after, and two and three weeks later.

\section{System Design}

\subsection{System overview}

Health misinformation persists in low-literacy communities not because correct information is absent but because the channels that carry accurate information do not reach them: accurate information comes in written form, and the community health workers who deliver advice in person are few, overstretched, and share medical advice with little training \cite{seelam2024fact, 10.1145/1753326.1753610}.
To fill this gap, we propose culturally adaptive AI videos of healthcare providers who look like women from the community, speak in their language, and a pipeline that produces them and can scale to multiple other demographies.

\begin{figure*}
  \centering
  \includegraphics[width=\textwidth]{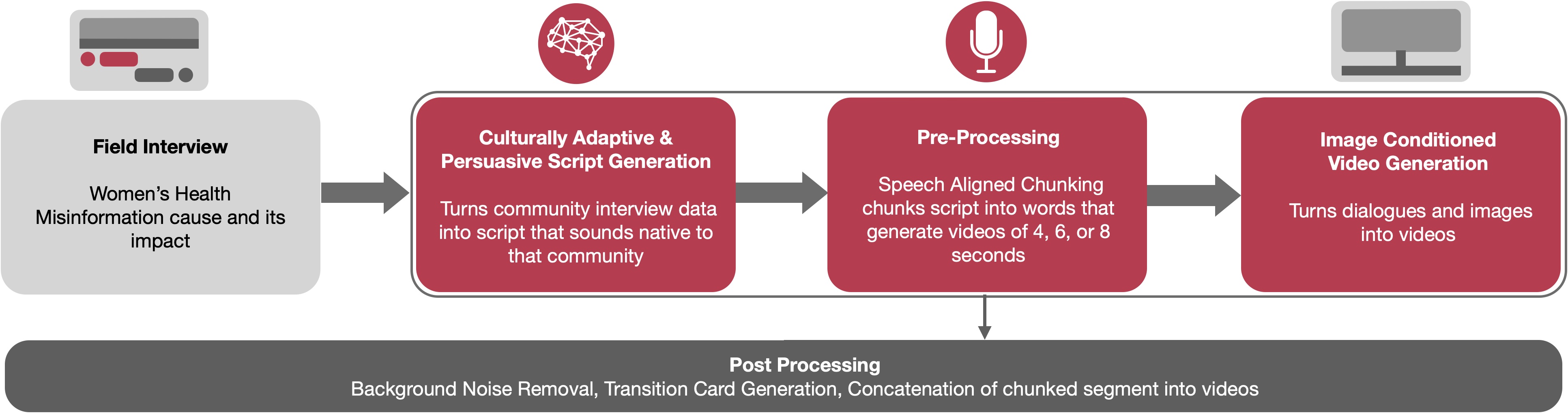}
  \caption{The five-stage pipeline. Stages (i)--(iii) are shared across conditions and produce a
    single script; only stage (iv) is run twice, conditioned on a different reference photograph.
    This ordering is what allows the two videos to be identical in what is said and to differ
    only in who appears to say it.}
  \label{fig:system_design}
  \Description{Pipeline diagram with five stages: field interviews producing a misinformation
    dataset, persuasive script generation, speech-aligned chunking, image-conditioned video
    generation run separately for the culturally adaptive and culturally neutral presenters, and
    post-processing and assembly into two complete videos.}
\end{figure*}

As described in Figure \ref{fig:system_design}, the system has five stages: (i) Field interviews that result in a dataset of prevalent women's health misinformation, (ii) Culturally adaptive persuasive script generation, (iii) Speech-aligned chunking, (iv) Image-conditioned video generation, and (v) post-processing.

\subsection{Field interviews and the misinformation dataset}
\label{sec:misinformation_data}

Women's health, and reproductive health in particular, remains markedly understudied relative to its need \cite{tuli2019sa, chowdhury2025literaturerewadiscussingreproductive}, and work around misinformation about women's health is limited. To the best of our knowledge, no dataset of health misinformation in resource-constrained communities exists, so we built one.

We conducted a formative field study with twelve health care providers from the same community, comprising doctors, pharmacists, and Accredited Social Health Activists (ASHAs). This study was approved by our Institutional Review Board (Protocol \#E-7161), and all participants provided informed consent. Each contributed through a 30--40 minute semi-structured interview or a written questionnaire on the same 21-item survey (Refer to Appendix~\ref{app:interview} and Fig .~\ref {fig:field-study-questions}). The survey moves from prevalence (which false beliefs providers encounter among women) to harms (which they consider most damaging, and concrete examples where a belief led to a bad outcome), then to persistence (which beliefs survive health education, and what makes them resistant), and finally to mitigation practice: how providers explain medical concepts to patients with no formal education, which analogies work, and whom the community actually trusts for health information. The last two groups of items shaped our second stage of the pipeline, which generates culturally adaptive and persuasive scripts to persuade women towards the correct health information.

\begin{figure*}
    \centering
    \includegraphics[width=1\textwidth]{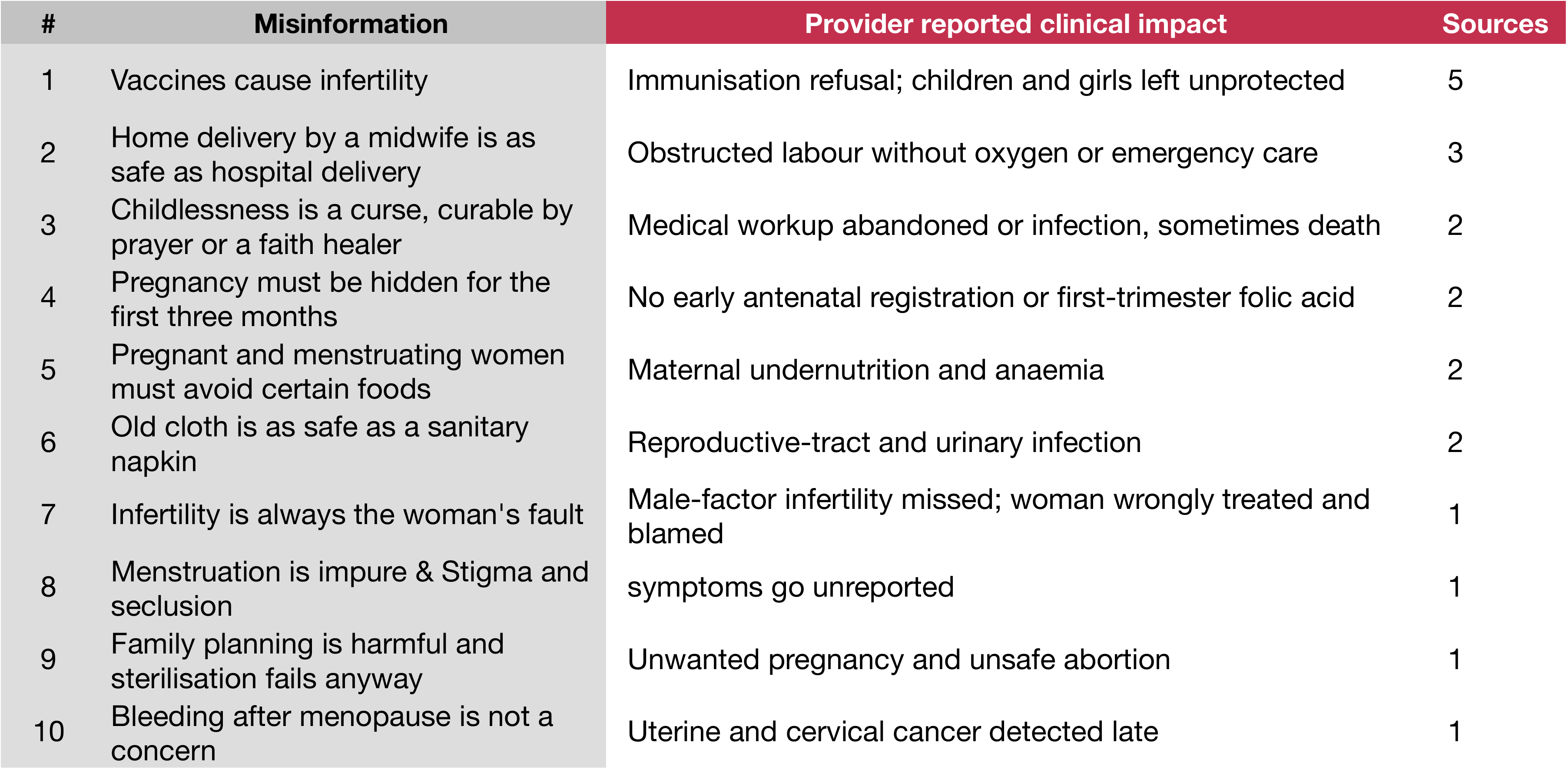}
    
    \caption{The ten women's health beliefs addressed by the video, elicited from twelve providers and ranked by the number of independent providers reporting each. Clinical impacts are as stated by providers; the released dataset pairs each with a verbatim de-identified quotation.}
    \Description{A table listing ten women's health misinformation beliefs, each paired with its provider-reported clinical impact and the number of providers who reported it. The list is ranked from most to least reported. Vaccines causing infertility is reported most often, followed by home delivery being as safe as hospital delivery. Other beliefs include childlessness as a curse, hiding pregnancy, food restrictions, unsafe menstrual cloth, blaming women for infertility, menstrual impurity, distrust of family planning, and dismissing postmenopausal bleeding. Impacts range from immunization refusal and obstructed labor to infections, anemia, unsafe abortion, and late cancer detection.}
    \label{fig:misinformation-evidence}
\end{figure*}

We transcribed and translated interviews with Saaras V3 \cite{sarvam2026saaras}, a speech-to-text model for code-mixed Hindi--English. We extracted every false or clinically contradicted belief, leading to a list of 43 distinct misinformation items, each annotated with the set of providers supporting it and ranked by that count. Filtering to women's health left 15 beliefs, of which 10 were retained for this study. These span misinformation about pregnancy, menstruation, infertility, contraception, vaccination, and cancer. Ten were carried into the video. Figure \ref{fig:misinformation-evidence} presents a complete list of misinformation, and Figure \ref{fig:misinf-imp} includes some of it, supported by a verbatim quotation retained in the dataset we will release. 

\subsection{Culturally adaptive \& persuasive script Generation}

Research on persuasion for conspiracies \cite{costello2024durably} or multimodal fake information \cite{rani2026culturescoreevaluatingculturalfaithfulness} has shown that persuasive dialogues lead to a reduction in false beliefs.
Conditioned on the dataset from Section~\ref{sec:misinformation_data}, each script follows a fixed three-part structure: the belief is first voiced in reported speech, as a woman would hear it in her own neighborhood (\emph{``you must have heard people say\ldots''}); then explicitly rejected; then replaced with a persuasive explanation and concrete guidance, closing on the trusted healthcare the interviews identified. Prompts are shared in Appendix~\ref{app:prompts}. Initial scripts averaged 190 words per misinformation. These were condensed to roughly 60 seconds of speech ($\sim$ 122 words) and revised to adapt to low-literacy listeners. Two healthcare providers then reviewed every script for comprehensibility in the study communities to validate the replacement of clinical vocabulary with words used locally—for menstruation, menopause, and the reproductive anatomy the scripts refer to. An example translated into English is present in Table \ref{tab:script-revision}.

\subsection{Speech-aligned chunking}
Video generation models generate short clips of fixed duration of 4,6 or 8 seconds, so a long video cannot be generated in one pass; it has to be assembled from many separately generated segments. Splitting on a fixed word count cuts sentences mid-clause, so the presenter appears to break off and restart at every seam, leaving words clipped at the end or the remainder padded with silence. We therefore split on linguistic boundaries rather than on counts. Sentences are packed greedily up to a 20-word budget; a sentence exceeding that budget is split at clause boundaries into near-equal pieces rather than greedily filled, which avoids a long piece followed by a stub; and any fragment under five words is merged into whichever neighbor can absorb it. Each chunk is then assigned a clip duration of 4, 6, or 8 seconds from its word count, calibrated to an observed Hindi speech rate of 2.6 words per second, so that short lines do not generate clips padded with silence. Applied to the ten misinformation items, this results in a chunked script that is used for generating video.

\subsection{Image-conditioned video generation}

Our study design requires two videos that differ only in the presenter's cultural identity. Generation was constrained accordingly so that the appearance dont drift between clips, and framing, pacing, and delivery must be identical across conditions. Each clip was generated using Veo~3.1 (\texttt{veo-3.1-generate-001}) \cite{GoogleDeepMind2025Veo3} in portrait 9:16.
The presenter images were produced with the Gemini image-to-image generation model from context photographs of women from the study communities, holding apparent age constant; the culturally adaptive condition depicts a community health worker in a rural Indian setting, and the culturally neutral condition depicts a community health worker in a neutral setting. In both of the videos, she speaks the same Hindi line, and clip duration, sequence, and runtime are identical by construction. Prompts can be found in Appendix section \ref{app:prompts}.

\subsection{Post-processing}
Generated videos occasionally carry residual music despite a negative prompt suppressing it, so each clip is passed through source separation so that residual music is suppressed, and only the vocal stem remains. Clips are concatenated per misinformation, and the ten misinformation segments are interleaved with transition cards announcing the misinformation about to be addressed, narrated with Gemini Flash TTS (voice\emph{Achernar}, \texttt{hi-IN}) \cite{GoogleDeepMind2025GeminiTTS}. 
Each video opens with a generated introduction in which the presenter greets the viewer and names herself \textit{Ganga Devi} in the culturally adaptive condition, \textit{Amy} in the culturally neutral condition, and both state that she is an artificial health assistant. The intervention therefore discloses its synthetic nature to every 
participant before any health content is delivered. The resulting segments are concatenated into a video of approximately 12 minutes.

\subsection{System evaluation}

We evaluate the generated videos on two axes: cultural faithfulness and \emph{equal} generation quality, which is the confound. We measure  (i) VideoScore \cite{he2024videoscorebuildingautomaticmetrics}, which predicts fine-grained human judgments of generated video along five dimensions. \emph{Visual quality} and \emph{temporal consistency} capture whether the presenter is rendered cleanly and stably across a long video. \emph{Text-to-video alignment} checks that each clip generates the line it was given. \emph{Factual consistency} checks that no clip departs from its script. \emph{Dynamic degree} measures motion and is reported as a conformance check rather than a quality score: our videos are deliberately locked-off static shots, so a low value is the intended behavior; (ii) CultureScore \cite{rani2026culturescoreevaluatingculturalfaithfulness} rates how faithfully generated video reflects a target culture's visual and social conventions. We used it to select the generator: Veo~3.1 produced video and synchronized audio aligning most closely with human cultural preferences among the models we compared. Table~\ref{tab:system-eval} reports both videos on all six measures. The comparison of interest is not that either video scores highly, but that they score \emph{alike} everywhere except CultureScore.

\begin{table*}[t]
\setlength{\tabcolsep}{6pt}
\centering
\begin{tabular}{lcccccc}
\toprule
 & & \multicolumn{5}{c}{\textbf{VideoScore}} \\
\cmidrule(lr){3-7}
\textbf{System} & \textbf{CultureScore} &
\begin{tabular}[c]{@{}c@{}}\textbf{Visual}\\\textbf{Quality}\end{tabular} &
\begin{tabular}[c]{@{}c@{}}\textbf{Temporal}\\\textbf{Consistency}\end{tabular} &
\begin{tabular}[c]{@{}c@{}}\textbf{Dynamic}\\\textbf{Degree}\end{tabular} &
\begin{tabular}[c]{@{}c@{}}\textbf{Text-to-Video}\\\textbf{Alignment}\end{tabular} &
\begin{tabular}[c]{@{}c@{}}\textbf{Factual}\\\textbf{Consistency}\end{tabular} \\ \midrule
Culturally adaptive & 80.0\% & 3.2 & 3.7 & 2.8 & 3.0 & 3.4 \\
Culturally neutral  & 10.8\% & 3.8 & 3.8 & 2.7 & 3.1 & 3.6 \\ \bottomrule
\end{tabular}
\caption{\textbf{Evaluation of the two generated videos.} Both videos were produced by the same pipeline from an identical script and differ only in the presenter's cultural identity. VideoScore rates five dimensions from 1 (lowest) to 4 (highest). Adaptation raised CultureScore from 10.8\% to 80.0\%, while both videos scored above the scale midpoint on every VideoScore dimension.}
\Description{A table comparing culturally adaptive and culturally neutral videos. CultureScore is 80.0 percent for adaptive and 10.8 percent for neutral. VideoScore ratings on a 1 to 4 scale.}
\label{tab:system-eval}
\end{table*}

\section{Study Design}

\subsection{Participants}
\label{sec:participants}
\begin{figure*}
    \centering
    \includegraphics[width=\textwidth]{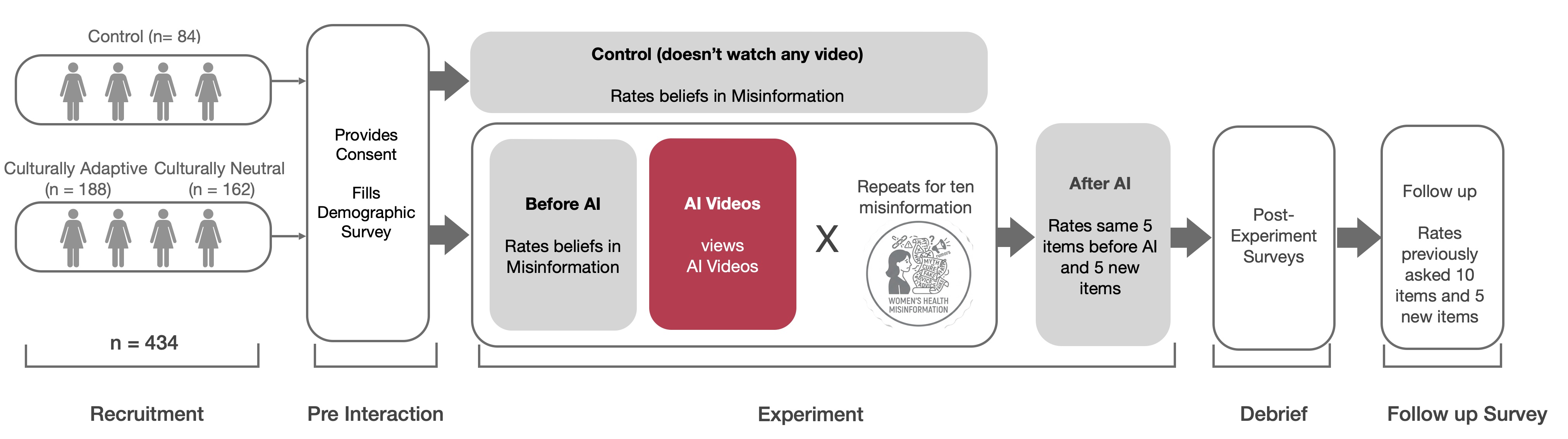}
    \caption{Longitudinal study design workflow showing the three-phase and three-condition experimental protocol. At Phase 1, participants rate five belief items, view a twelve-minute AI health video from a culturally adaptive presenter or a culturally neutral one delivering an identical script, and re-rate those five alongside five never asked before. control group is surveyed once only at phase 1. Participants return after 2 and 3 weeks, each session adding five previously unasked items. Every claim contributes two differently worded items in different sets, so each session measures belief on claims asked beforehand and on claims never raised — separating correction of elicited statements from learning, and immediate change from its retention.}
    \label{fig:study}
    \Description{Flowchart of the study across five stages: recruitment, pre-interaction, experiment, debrief, and follow-up. Of 434 participants, 84 in the control group only rate misinformation beliefs, while 188 culturally adaptive and 162 culturally neutral participants rate beliefs, watch AI videos, and re-rate the same five items plus five new ones. All proceed to post-experiment surveys, then a follow-up rating the ten earlier items and five new items.}
\end{figure*}

We recruited participants from the Mangolpuri district of North West Delhi, a region characterized by its predominantly rural landscape, limited access to tertiary healthcare facilities, and largely driven by daily wage labourer that significantly influence health-seeking behaviors and the persistence of misinformation \cite{chaplin2017infrastructure}. 
We conducted an eligibility screening with 800 women across Mangolpuri with the help of a local organization that runs a women's literacy center to educate low-literacy women regarding basic things like how to dial a number on a phone or read a bus number. 800 women rated their beliefs on five prevalent misinformation items reported by doctors in their community. Post-screening, 583 met the criterion of having a belief in at least one misinformation item and were randomly assigned to a culturally adaptive AI video condition, a culturally neutral AI video condition, or a control condition where participants did not receive any intervention. Of these, 467 completed the first phase (190 culturally adaptive, 170 culturally neutral, and 107 control), in which participants rated their beliefs, viewed their assigned video, and rated their beliefs again; control participants completed only the belief ratings, without viewing a video. 25 dropped out before completing all three sessions, and we excluded an additional 8 participants who failed more than two attention checks across all sessions. At two points in each session, embedded within the belief blocks, the interviewer recorded her own judgment of whether the participant appeared to be answering attentively; a participant was flagged if the interviewer judged her inattentive. A total of 434 women completed the entire study; 350 women in the two video conditions completed all three phases (baseline and follow-ups at two and three weeks; see Section \ref{sec:experimental_protocol}), and the control condition (n=84) was surveyed at baseline only. Figure \ref{fig:study} illustrates the study design.

\textbf{Demographics.} Out of 434 women whose data are analyzed for this study, 77\% were aged 25 to 54. Phone access is broad but shallow: three quarters own a mobile phone, yet only 56\% operate one unaided, 26\% need help, and 17\% cannot use one at all. Their fluency is concentrated in exactly the media the intervention resembles --- 68\% use WhatsApp daily and 62\% use YouTube. Two-fifths cannot use a phone without help; exposure was interviewer-led. Appendix~\ref{app:demographics} and Figure ~\ref{fig:demography} report the full profile.

Participants received 500 rupees in compensation for a combined total of approximately 50–60 minutes of contact across all three study sessions. The interviewer received 1000 rupees for every session. This is compliant with the minimum wage of 710 rupees per day for unskilled workers in this area \cite{delhi2026minwage}: 500 rupees for roughly an hour of participation is well above the statutory hourly rate. This study was approved by our Institutional Review Board at MIT (Protocol
\# E-7912), and all participants provided informed consent.

\subsection{Experimental protocol}
\label{sec:experimental_protocol}

Each participant completed three sessions over a four-week field period: a baseline session that ran for a week, a first follow-up after two weeks, and a second follow-up after three weeks. In phase 1, participants in both conditions first answered a five-item belief block administered before any exposure, then watched the twelve-minute video in a group setting, typically alongside other women at the same literacy center, then answered a ten-item belief block: the same five items again, plus five new items -- differently worded questions about the video's misinformation 
that the participant had not previously encountered. Belief ratings and the post-exposure interview were administered individually and privately at every phase. The control condition 
group completed the identical question blocks with no video shown between them. Examples of some items are present in Figure \ref{fig:question-item-example} and an extended list in Appendix 
section \ref{tab:item-schedule}.

  \begin{figure*}
    \centering
    \includegraphics[width=\textwidth]{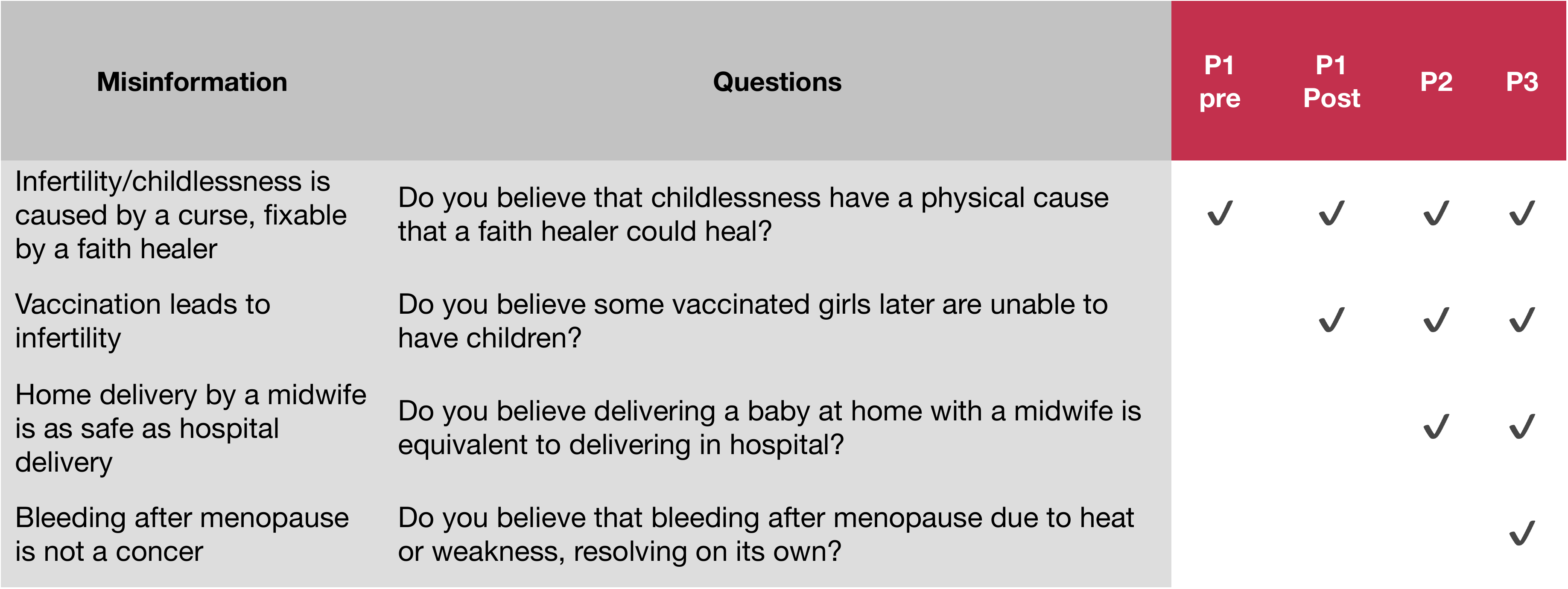}
    \caption{\textbf{Example questions across phases.} P1 pre and P1 post denote questions asked before and immediately after the intervention; P2 and P3 denote follow-ups at two and three weeks. Each phase repeats earlier questions and adds new ones.}
    \label{fig:question-item-example}
    \Description{A table with four rows showing when each misinformation question was asked. Columns are Misinformation, Questions, and four phase columns: P1 pre, P1 post, P2, and P3.}
\end{figure*}

At the first follow-up, after two weeks, all participants answered the previously asked 10 items from phase 1, plus five further items — differently worded questions about misinformation the video addressed but that had not yet been asked of this participant in either wording. At the second follow-up, participants answered another 5 similar new items alongside all prior items. Each of the ten misinformation addressed by the video contributes one item to two phases, so that a misinformation pre-tested in one phase is tested again, without warning, via its paired item in a different phase -- allowing us to separate belief correction of a specifically worded claim from learning transfer to a different phrasing of the same misinformation. This design allowed us to measure immediate belief change in misinformation, retention, and whether participants developed learning through the items they haven't interacted with in any of the phases.

\subsection{Data collection}

We collected beliefs, confidence, perceived societal norm, along with demographic and 
digital-access measures, and open-ended qualitative responses on 9 questions using MS Office forms filled out by interviewers. Belief ratings were collected for each misinformation item on a five-point scale. Each item is keyed $d \in \{-1, +1\}$ according to whether agreement or disagreement is the correct response (Figure~\ref{fig:question-item-example} has some examples and Table~\ref{tab:item-schedule} in the appendix contains the complete set). For a five-point response $x \in \{1, \ldots, 5\}$, we recenter the scale around its midpoint and rescale it to 
run from $-1$ to $+1$: the per-item score is $(x - 3) \times d / 2$, ranging from $-1$ (agreement with the misinformation) to $+1$ (agreement with correct information); \textit{don't know} responses are scored $0$. Ten items are keyed so that agreement is correct and ten so that disagreement is correct, to mitigate acquiescence bias -- a tendency to agree with statements regardless of content -- which is common in field studies in low-literacy areas \cite{narayan1996education}. A self-reported confidence rating on a three-point scale (full confidence, some confidence, no confidence at all) immediately follows each belief rating. For each misinformation item, we also ask a parallel question about the participant's community; for example, alongside ``Do you believe that using family planning methods such as contraceptive pills is safe for women's health?'', we further ask whether women in her community believe this, which lets us separate her personal belief from her perception of the social norm around it. Two items per session, embedded within the belief blocks, were captured in which interviewers were asked to confirm whether the participant was answering attentively.

\section{Analysis}
\label{sec:analysis}

To analyze the effect of culturally adaptive and neutral AI video interventions,
we ask three research questions:
\begin{enumerate}
    \item RQ1: Does a culturally adaptive video change belief more than a culturally neutral one within a single session, and does either change belief relative to a control group? 
    \item RQ2: Among the misinformation items the video addressed, does correction extend to a differently worded question about the same claim that the participant was never asked to rate?
    \item RQ3: Does the corrected belief persist two and three weeks later post-intervention?
\end{enumerate}
To answer RQ1, we measure \textbf{immediate belief change}, which is the change in belief on the pre-tested items from before to immediately after viewing. This is the standard outcome in single-exposure correction studies, and the only one for which we hold a pre-measurement of the same items in the same participants. 
To answer RQ2, we measure \textbf{learning}: for each of the ten misinformation items, the video is tested with two differently worded questions (Table~\ref{tab:item-schedule}), and each participant is asked only one wording of a given item per phase. Learning is the belief score on the wording a participant has not yet answered, compared against the same wording answered by the no-video control group. Because we hold no pre-video measurement for this specific wording in this specific participant, learning is inferred from a between-group contrast against control, not from a within-participant change, and it indicates whether correction generalizes across the phrasing of a claim the participant was never directly asked, rather than across claims. To answer RQ3, we measure \textbf{retention}, whether the immediate belief change is retained in the follow-up phases after 2 weeks and 3 weeks.

\subsection{Measure}
\label{sec:measure}

Each item has a correct direction $d \in \{-1, +1\}$: $d = +1$ if disagreement with the item is the correct response, and $d = -1$ if agreement is (Table~\ref{tab:item-schedule}). For a five-point response $x \in \{1, \ldots, 5\}$, we recenter the scale around its midpoint and rescale it to run from $-1$ to $+1$: the per-item score is $(x - 3) \times d / 2$, ranging from $-1$ (confidently endorsing the misinformation) to $+1$ (confidently rejecting it); \textit{don't know} is a neutral position and scores $0$. A participant's \textbf{belief score} on a set of items is the mean of her per-item scores. Five of the ten claims are worded so that agreement is correct and five so that disagreement is correct, to mitigate acquiescence bias; a tendency to agree with statements regardless of content, which is common in field studies in low-literacy areas \cite{narayan1996education}.

Because the score is bounded at $+1$, the same raw difference is worth more on some item sets than on others: a set on which untreated women already score $+0.23$ leaves less room to move than one on which they score $+0.04$. Where we quote a percentage, it is the condition's gain over the no-video control, expressed as a share of the distance remaining between the control's 
score and the ceiling of $+1$ (full rejection of the misinformation):
\[
\frac{\text{belief score}_\text{condition} - \text{belief score}_\text{control}}{1 - \text{belief score}_\text{control}}.
\]

For example, on the pre-tested items, the no-video control scored $+0.039$ and the culturally adaptive condition scored $+0.324$: a gain of $+0.285$. The remaining distance from control's score to the ceiling is $1 - 0.039 = 0.961$, and $0.285 / 0.961 \approx 30\%$, so we say the 
culturally adaptive condition closes 30\% of the gap to full rejection on these items. On the new items introduced after viewing, the no-video control scored $+0.233$ -- a higher starting point, leaving less room to move -- and the culturally adaptive condition scored $+0.123$ higher than control; here $0.123 / (1 - 0.233) \approx 16\%$.
Expressing both as a share of the room available lets us compare correction on misinformation a participant was asked about beforehand (30\%) with correction on misinformation she was not asked about (16\%), even though the two item sets have different ceilings. It is only a descriptive rescaling, however: we report it alongside the raw belief-score difference in every case, not in its place, and all statistical tests are performed 
on the raw, unscaled belief score rather than on this percentage.

\subsection{Models}

\textbf{RQ1.} 
We fit an ANCOVA \cite{vickers2001analysing} of post-video belief on condition, adjusting for pre-video belief:
\\
$\text{belief score}_\text{post} \sim \text{condition} + \text{belief score}_\text{pre}$ \\ 
We report the between-condition coefficient, its 95\% CI, and $d_\text{adj}$, the coefficient over the residual standard deviation. Adjustment is required because the conditions differ at baseline (Section~\ref{sec:analytic-sample}); we report no unadjusted between-condition contrast. Each condition is compared to the control group where there was no video intervention.

\textbf{RQ2.} We compare each condition's belief on the new items in the phase 1 set against control on the same set, and set that against the corresponding pre-tested-set contrast. 
We also compute each participant's own between-set difference and compare it to the control group. 
At phases 2 and 3, the comparison runs between conditions only, as control has no follow-up measurement. 
We report the minimum detectable effect so that a null is interpretable. Because a non-significant result can reflect either no effect or insufficient power, we report the smallest effect each comparison could reliably detect. At 80\% power, that is 0.147 belief score (d = 0.37) for the culturally adaptive condition against control, 0.126 (d = 0.38) for the culturally neutral one, and 0.111 (d = 0.30) between conditions. Differences smaller than these cannot be ruled out by these data.

\textbf{RQ3.} We test each condition's post-video belief score against its own belief score at each follow-up, paired within participants, reporting the confidence interval on the change so the largest decline consistent with the data is explicit. 
We then regress follow-up belief on pre-video and within-session gain: $\text{belief}_\text{followup} \sim \text{belief}_\text{pre} + \text{gain}$, testing whether participants who moved most during the session remain higher weeks later. This requires no untreated comparison group and is robust to regression to the mean, but is attenuated by measurement error in the gain term, so we treat it as a lower bound.

\section{Quantitative Results}

For the research questions, we ask in section \ref{sec:analysis} to measure immediate belief change, learning, and retention; we report three outcomes in the order the design produces them: the immediate change in belief across a single session, the extent to which that change appears on misinformation the participant was never asked about beforehand, and the belief score at follow-up. 
All scores are bounded at $-1$ (full agreement with misinformation) and $+1$ (full disagreement), with $0$ marking both the scale midpoint and the \emph{neutral} response.

\subsection{Analytic sample}
\label{sec:analytic-sample}

Of 467 women enrolled, 434 met the inclusion criteria and completed every session they were scheduled for: 188 culturally adaptive, 162 culturally neutral, and 84 no-video control. The control figure reflects successive attrition: 107 women were enrolled, 84 completed a usable baseline interview. By design, control participants were surveyed only once, at baseline; they did not return for the phase 2 or phase 3 follow-ups, since there was no intervention whose immediate or delayed effect needed to be tracked in this group. The conditions were not balanced at baseline. Pre-video scores on the pre-tested set were higher in the culturally adaptive condition ($+0.203$ vs $+0.135$; $\Delta = +0.069$, $t = 2.11$, $p = .036$). Randomization was performed individually, i.e., each woman was independently assigned to a condition rather than being assigned in a group, and stratified by literacy center, so that each condition received a proportional share of participants from every center. All between-condition tests below adjust for the pre-video score.

\subsection{Immediate belief change}

\begin{figure*}[t]
  \centering
  \includegraphics[width=2.1\columnwidth]{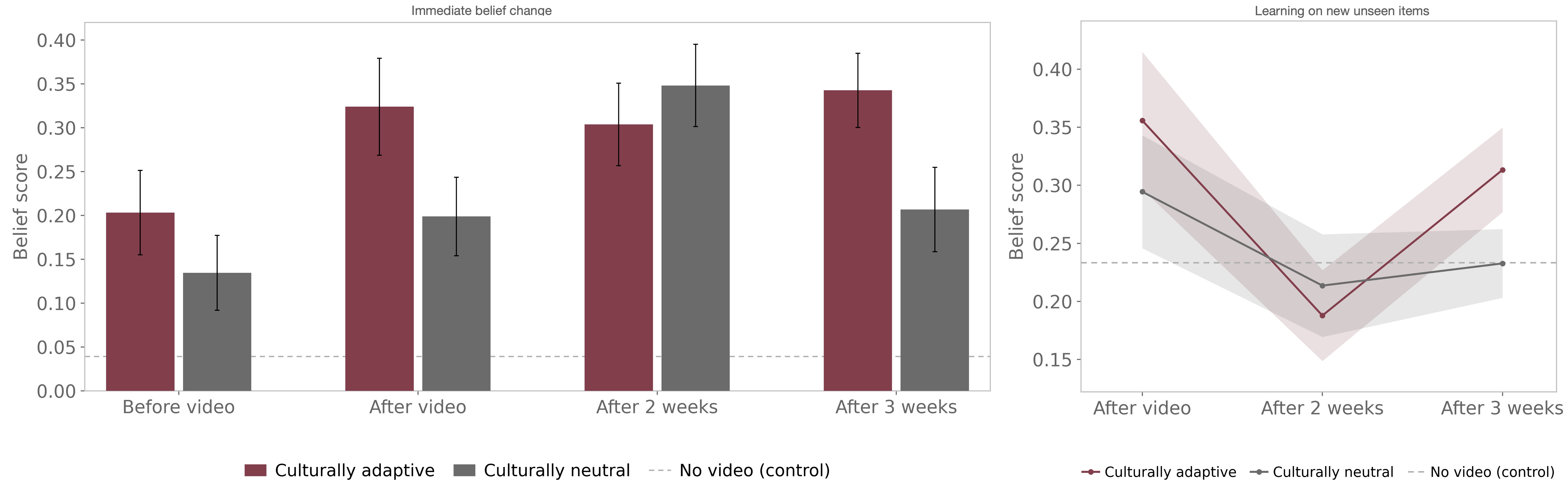}
  \caption{Belief score across conditions and sessions. Higher values indicate stronger rejection of the misinformation; the belief score runs from $-1$ (agreement with the misinformation) to $+1$ (disagreement with the misinformation), and one step on the five-point response scale moves a participant's score by $0.1$. Left: the five questions asked before viewing and repeated at every session. Both conditions rise within the session, the culturally adaptive condition roughly twice as far ($+0.121$ against $+0.064$), and neither loses ground by three weeks. The dashed line is the no-video control, measured once. Right: the five questions introduced at each session, never previously asked of that participant. Each session uses a different set, so points are comparable between conditions within a session but not across sessions; the control marker spans the first session only. Error bars and shaded bands are 95\% confidence intervals. ($N = 434$).}
  \Description{Two side-by-side charts of belief score, where higher means stronger rejection of misinformation. Left, a bar chart of repeated items compares culturally adaptive and culturally neutral conditions before the video, after the video, and at two and three weeks, with a dashed line for the no-video control. Both conditions rise after the video, with the adaptive condition rising more; the neutral condition briefly overtakes it at two weeks. Right, a line chart with shaded confidence bands shows scores on newly introduced items at each session. The adaptive condition leads after the video, dips below the neutral condition at two weeks, and leads again at three weeks.}
  \label{fig:belief-change}
\end{figure*}

Both video conditions shifted belief away from the misinformation over the course of the phase 1 session, which included a single visit in which participants rated their beliefs, watched the video, and rated their beliefs again. The culturally adaptive condition rose from $+0.203$ to $+0.324$, a gain of $+0.121$ (95\% CI $[0.064, 0.177]$, $t(187) = 4.23$, $p < .001$, $d_z = 0.31$). The culturally neutral condition rose from $+0.135$ to $+0.199$, a gain of $+0.064$ (95\% CI $[0.009, 0.119]$, $t(161) = 2.29$, $p = .023$, $d_z = 0.18$). Cultural adaptation roughly doubled that effect. Adjusting for pre-video score, the culturally adaptive condition ended the session $0.099$ higher (95\% CI $[0.031, 0.168]$, $p = .005$, $n = 350$, $d_\text{adj} = 0.31$). Because the two videos were generated from one script, this difference is attributable to the presenter's cultural identity alone. Both conditions also ended well above the no-video control, which scored $+0.039$ on this set without any intervention: $+0.285$ higher for the culturally adaptive condition ($t = 6.28$, $p < .001$, $d = 0.80$) and $+0.160$ higher for the culturally neutral condition ($t = 3.77$, $p < .001$, $d = 0.52$). Relative to the control's score of $+0.039$, the culturally adaptive condition closes 30\% of the remaining distance to full rejection ($+1$), and the culturally neutral condition closes 17\% (Section~\ref{sec:measure}). Figure \ref{fig:belief-change} illustrates the finding.

\subsection{Learning}
\label{sec:learning}

Correction transferred weakly to claims participants had never rated. On the new items introduced after the video, the culturally adaptive condition remained above control ($+0.123$, $t = 2.44$, $p = .016$, $d = 0.31$), while the culturally neutral condition did not ($+0.061$, $t = 1.29$, $p = .20$, $d = 0.18$). Women in the control group, who saw no video, scored $+0.233$ on this set. 
Relative to that baseline, the culturally adaptive condition closes 16\% of the remaining distance to full rejection, and the culturally neutral condition closes 8\%, compared with 30\% 
and 17\% on the claims raised before viewing; roughly half the correction, in each condition, relative to claims the participant had already been asked about (Figure~\ref{fig:belief-change}).

Both of those comparisons are between groups that already differed before viewing (Section~\ref{sec:analytic-sample}). We therefore also compared each participant's own between-set difference against the control group's, which uses each woman as her own anchor and the control's between-set gap as the item-difficulty baseline. Measured from her pre-video level on the pre-tested set, the new items scored $+0.153$ higher in the culturally adaptive condition and $+0.160$ higher in the culturally neutral condition, against $+0.194$ in control; neither differs from control ($-0.041$, $t = -0.74$, $p = .46$; $-0.034$, $t = -0.63$, $p = .53$). Measured this way, the new items show no advantage for either condition over control, so this analysis finds no evidence of learning beyond what the item sets' difficulty alone would predict.

\subsection{Retention}

Neither condition reduced their accuracy. Across the ten repeated items, the culturally adaptive condition was unchanged from after video to 2 weeks later ($-0.034$, $p = .25$) and from 2 weeks later to three weeks later ($+0.036$, $p = .25$). The culturally neutral condition rose at 2 weeks later ($+0.076$, $p < .001$) and returned to its after-video level by week 3 ($-0.009$, $p = .74$).
Against the before-video baseline, both retained a gain three weeks later: $+0.139$ for the culturally adaptive condition (95\% CI $[0.081, 0.198]$, $p < .001$, $d_z = 0.34$) and $+0.072$ for the culturally neutral condition (95\% CI $[0.008, 0.136]$, $p = .027$, $d_z = 0.18$).

The between-condition difference is present at phase 1, absent at phase 2, and largest at phase 3. Adjusted for baseline, the culturally adaptive advantage was $+0.099$ at phase 1 ($p = .005$, $d = 0.31$), $-0.053$ at phase 2 ($p = .12$), and $+0.129$ at phase~3 ($p < .001$, $d = 0.43$, $n = 350$). Phase 2 is also the session at which the culturally neutral condition records an isolated rise that reverses within a week to phase 3. Control has no follow-up measurement, so that rise cannot be separated from retesting,
maturation, or between-session exposure, and we draw no conclusion from it.

\begin{figure*}[t]
  \centering
  \includegraphics[width=2.1\columnwidth]{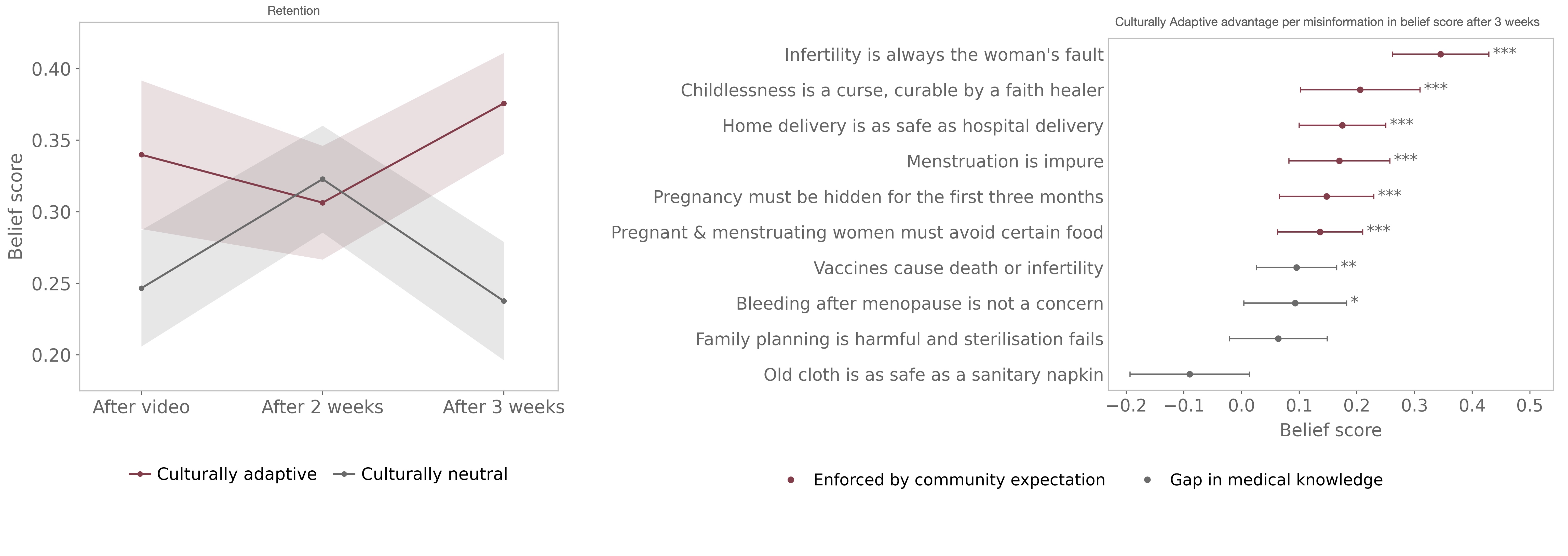}
  \caption{\textbf{Retention and how beliefs about misinformation changed.} \textbf{Left:} the ten questions repeated at every session, in belief score ($-1$ to $+1$, higher indicates stronger rejection of the misinformation). Neither condition lost ground between the session and three weeks: the culturally adaptive condition was unchanged ($-0.034$ then $+0.036$), while the culturally neutral one rose at two weeks before returning to its Phase~1 level ($-0.009$). The culturally adaptive advantage was still intact at three weeks and at its widest ($+0.129$, $p<.001$, $d=0.43$). Shaded bands are 95\% confidence intervals. \textbf{Right:} the culturally adaptive advantage at three weeks for each of the ten pieces of misinformation, with both of its questions answered, coloured by how the belief is held. Adaptation acted almost entirely on claims sustained by community expectation ($+0.192$) rather than on gaps in medical knowledge ($+0.041$); the interaction is $+0.156$ ($p<.001$). Bars are 95\% CIs; \textsuperscript{***}$p<.001$, \textsuperscript{**}$p<.01$, \textsuperscript{*}$p<.05$. }
  \Description{Two charts. Left, a line chart titled Retention, shows belief scores on repeated items after the video, at two weeks, and at three weeks, with shaded confidence bands. The culturally adaptive condition stays roughly level and ends highest at three weeks, while the culturally neutral condition rises at two weeks and falls back by three weeks, leaving a wide gap between the two. Right, a dot plot with confidence intervals shows the culturally adaptive advantage at three weeks for each of ten misinformation claims, sorted from largest to smallest. Claims enforced by community expectation, such as infertility being the woman's fault, childlessness being a curse, and pregnancy needing to be hidden, are shown in dark red and have the largest, statistically significant advantages. Claims reflecting gaps in medical knowledge, such as vaccines causing infertility and old cloth being as safe as sanitary napkins, are shown in grey.}
\label{fig:retention-mythtype}

\end{figure*}

Retention here is a group-level property. The carry-through model is significant after 2 weeks in both conditions ($+0.176$, $p = .009$; $+0.203$, $p = .016$) but not after 3 weeks ($+0.055$, $p = .37$; $+0.098$, $p = .26$). Test--retest reliability across sessions is $r = .24$, so these coefficients are attenuated and should be read as lower bounds. Figure \ref{fig:retention-mythtype} illustrates retention.

\subsection{Which beliefs changed}
\label{sec:mythtype}

The ten misinformation items are divided by how they are held. Six are sustained by family and community norms: who is blamed for childlessness, what is impure, what a pregnant woman must conceal or avoid eating, where she should give birth. Four are gaps in medical knowledge: what a vaccine does, whether contraception is safe, what bleeding after menopause signifies, whether cloth is as safe as a sanitary napkin. The grouping was made by the research team and endorsed by our social-impact partners.

Cultural adaptation acted mainly on the first group (Figure~\ref{fig:retention-mythtype}). At Phase~3 (after 3 weeks), adjusting for baseline, the culturally adaptive advantage was $+0.192$ on norm-based claims (95\% CI $[0.141, 0.243]$, $p < .001$, $d = 0.80$) and $+0.041$ on knowledge-based claims (95\% CI $[-0.007, 0.090]$, $p = .10$, $d = 0.18$). The interaction is $+0.156$
(95\% CI $[0.104, 0.209]$, $p < .001$, $d = 0.64$) and survives adjustment for the same contrast at baseline, which contributes nothing (coefficient $-0.001$). All six norm-based claims reach $p < .001$, and each has a larger effect than any knowledge-based claim ($d \geq 0.38$ against $d \leq 0.29$), though two knowledge-based claims are themselves significant. The largest single effect is on \emph{infertility is always the woman's fault} ($+0.345$, $d = 0.88$); the only reversal is on \emph{old cloth is as
safe as a sanitary napkin} ($-0.090$, $p = .09$). 

This split holds for belief measured at phase 3, after every misinformation had been rated at least once; it does not appear as a participant's first exposure to the misinformation, so it is not an account of learning. However, the property that made a belief movable here was not how false it was but how socially it was enforced: claims a woman is expected to hold cost her standing to abandon, and those are the ones cultural adaptation shifted. Correcting such a belief requires a competing source with standing in her own community, which a matched presenter supplies and a neutral one does not; where the barrier was only an information gap, the neutral presenter was already enough. We would therefore expect cultural adaptability to generalize to domains where practice is policed by community expectation --- vaccination, contraception, nutrition, mental health --- and to add little where the deficit is purely informational.

\section{Qualitative Analysis}
\label{sec:qual}

After the intervention, 434 participants were interviewed for 15 minutes individually on nine open qualitative questions: what the video was about and her reaction to it; who she thought it was made for; whether she had encountered this specific misinformation before the study, and if so, where from; what someone who firmly believed it would say; what felt familiar; how the group watching the video together seemed to receive it, and whether her own reaction might have differed if she had watched alone; whether she would pass it on, and to whom not; where these topics arise in her life; and what she thought the study was about.

We then conducted a thematic analysis to identify overlapping positions among participants. Inductive coding \cite{boyatzis1998transforming} was used to discover salient themes: two of the paper's co-authors coded the interview transcripts, developing codes by reading the responses, refining them against further reading, and applying them to the full corpus, with emerging themes added until reaching saturation \cite{creswell2017research}. Our thematic analysis revealed seven recurring positions from which women described receiving the video's corrective message. The quotes were translated into English while preserving the meaning. These themes are non-exclusive: participants could contribute to multiple patterns, so category counts do not sum to the full sample.

\begin{enumerate}

\item \textbf{Informational vs. Transformational Change (n = 177, 51\%).} Kegan distinguishes informational learning, which changes what a person knows, from transformational learning, which changes how she knows \cite{kegan2009form}. Both appear here, and they are not equally consequential. Of these participants, 128 describe only a fact arriving (P4, P7, P10, P14, P17, P18, P19,
P174); 49 describe the grounds on which they accept a claim shifting underneath them (P5, P27, P37, P50, P51, P53, P56, P61, P63).

``\textit{I found out about family planning methods, which I didn't know much about before. My reaction didn't change; it stayed the same right through the video}'' (P174).

``\textit{Before, I thought that what the old people say is from their own experience. But after watching the video, I felt the doctor should also be listened to, and one should not accept every old saying without understanding it}'' (P56).

The two framings track belief. Participants who described only a fact arriving gained almost nothing over the session, while those who described the grounds shifting gained several times as much, a difference of $+0.174$ ($p = .009$) that survives adjustment for how much was written, for baseline belief and for condition. By three weeks the two groups no longer
differed. Changing how a woman knows bought a substantially larger immediate correction, not a more durable one.

\item \textbf{Culturally Individuated Presenter (n = 173, 49\%).} 
When asked directly whether the woman on screen was a real person or was made by a computer, half the sample judged her as real. Many go further, referring to her by name or kin title while describing what they learned, treating her as a source rather than a channel (P6, P11, P28, P117, P178). Both videos name the presenter in the opening introduction: Ganga Devi in the culturally adaptive condition, Amy in the culturally neutral condition; but no participant who saw the culturally neutral presenter uses her name anywhere in the corpus, even though it was equally available to them.

``\textit{Some things we did not know before; Ganga didi was explaining them in the video}'' (P11).

\item \textbf{Household Origins of Misinformation (n = 189, 54\%).} These participants locate the origin of misinformation in a named person, household interaction, or particular occasion. For instance, a sister-in-law's pregnancy, an uncle paying a faith healer, an aunt with no one to ask (P1, P5, P11, P20, P162, P253, P347).

``\textit{My own uncle; his wife wasn't having a son, so they had an exorcism done and gave a lot of money to a faith healer}'' (P253).

``\textit{My aunt was pregnant and did not have full information, and because of that she lost the baby}'' (P162).

\item \textbf{Attributional Distancing (n = 58, 17\%).} Asked to imagine how someone who firmly believed the misinformation would respond to the video, these participants readily supplied a detailed rebuttal, but consistently attributed them to a third person rather than endorsing them themselves (P41, P42, P100, P103, P158, P343). The imagined objection was rarely factual; instead, participants framed resistance as a violation of community norms or tradition. 

``\textit{They will say this is a lie; you are spoiling our tradition. They won't believe childlessness is not a curse}'' (P158).

\item \textbf{Social Hesitancy (n = 59, 17\%).} Most participants say they would pass the video on. These participants, however, name a specific social cost to doing so: not the effort of sharing it, but the risk to her standing in the family if she does (P11, P50, P55, P57, P83, P346).

``\textit{There will be some hesitation in showing it to my husband and mother-in-law; they may feel I am acting too clever. But I will still try}'' (P346).

The correction moves through the same family relationships that brought the misinformation. What makes it costly is the direction of travel: to tell a husband or a mother-in-law is to contradict someone senior, and the cost these women name is not the effort but being seen as acting too clever.

\item \textbf{Authority Substitution (n = 176, 50\%).} Where these participants describe a belief giving way, the mechanism is one source replacing another --- the doctor or the ASHA (Accredited Social Health Activist, a government-appointed community health worker) taking the position the elder previously held. (P4, P11, P50, P55, P57, P83).

``\textit{The hardest thing for her will be accepting that what she got from the elders of her house is wrong. But after watching the video, a doubt may enter her mind, and she may think of confirming with a doctor}'' (P55). 

\item \textbf{Someone like Her, Not More Information (n = 81, 23\%).} We asked where the misinformation topic comes up in her life. These participants do not describe missing information; they describe missing \emph{someone} — available, from her own cultural identity, who understands her language, who matters at the moment, and who will not scold her for asking (P7, P50, P71, P160, P190, P346, P347, P359).

``\textit{I was looking for someone of my own; an ASHA or ANM didi who would explain in our language, and \textbf{not scold}. You have to go to the doctor, who doesn't give you time}'' (P347).

``\textit{When you can't be certain what is true, you feel there should be someone of your own that you could ask}'' (P359).\\ 

\end{enumerate}

Across the seven themes, the interviews converge on the same conclusion the quantitative results reach but by a different route. What separated the two conditions was that in one of them the presenter was perceived as a person who spoke rather than a video that played. And what accompanied the largest belief change was not more information received but a shift in the grounds on which a claim is accepted, a shift that brought a larger immediate correction. Cultural adaptation and transformational framing act the same way: both change how much belief change results from a misinformation mitigation effort.

The themes also locate what the intervention is competing with. Misinformation arrives from a mother-in-law, sister-in-law, a woman in the next house; when a correction is refused, the refusal is framed as disloyalty rather than as a factual disagreement, and belief gives way not to a better argument but to a competing person with social standing. That is why a culturally  adaptive AI is a plausible instrument here, as it offered our participants a recognizable and relatable authority figure with credible information.

\section{Discussion}

\subsection{Changing belief vs. being perceived as relatable}

Our findings separate two things the similarity-attraction account treats as one. Prior work on agent and avatar similarity finds that demographic matching reliably changes how people regard a messenger without reliably changing what they learn \cite{baylor2004pedagogical, wang2026similar, lee2026health}. We find the opposite pattern, and one closer to the field experiments on messenger identity in which behavior rather than rated similarity moved with the messenger \cite{alsan2019diversity, alsan2024experimental}: Belief in the misinformation was reduced more with a culturally adaptive presenter ($d=0.31$ immediately, $d=0.43$ at three weeks), while participants in that condition did \emph{not} report feeling more addressed than participants who watched a culturally neutral presenter deliver identical words. No participant described the presenter as foreign, and none indicated awareness that two versions existed.

An effect invisible to the people it acts on is awkward for the mechanism usually proposed for it. If matching worked by making a viewer feel spoken to, the women who moved most should have been the women who reported being spoken to. They were not. What distinguished the conditions was individuation. Eleven women referred to the
culturally adaptive presenter by name or kin title, \textit{Sahayika didi}, the term
they use for an ASHA worker, and none named the neutral presenter, who was credited as
``the video'' or as the interviewer. In the categorization--individuation account of
face perception, in-group faces are encoded as individuals and out-group faces as
members of a category \cite{hugenberg2010categorization}, the same continuum from
category-based to individuated impressions that governs person perception generally
\cite{fiske1990continuum}. Here the matched presenter was encoded as a person with
standing, the unmatched one as a medium. Individuation is a different construct from
rated similarity or perceived tailoring, and it was the only one that tracked belief
change.

\subsection{Cultural adaptation moves norms rather than knowledge gaps}
\label{sec:discussion-norms}

The advantage is not spread evenly across the misinformation. Grouping the ten claims by how they are held, the culturally adaptive advantage at three weeks was $+0.192$ on claims sustained by family and community expectation and $+0.041$ on claims that are gaps in medical knowledge, an interaction of $+0.156$ ($p<.001$). Every norm-based claim separates; the largest single effect is on \emph{infertility is always the woman's fault}.

The reading follows from the distinction set up in Section~\ref{sec:norms-vs-knowledge}. A claim held because a mother-in-law says so is a social norm in Bicchieri's sense \cite{bicchieri2017norms}: it is not corrected by information alone, because information was never what sustained it, and what is needed is a competing source with standing in the same social world, or evidence that the reference group's expectations are not what they seemed \cite{bursztyn2020misperceived}. A claim held because nobody has explained what a vaccine does is an information deficit, and any credible source closes it --- here both conditions did, and cultural adaptation added little. The result also refines the surface- versus deep-structure account of tailoring \cite{resnicow1999cultural, kreuter2003achieving}. Our manipulation was pure surface structure, which that account expects to buy attention rather than persuasion; yet it moved the norm-held beliefs, the ones deep-structure tailoring is meant for. A plausible reconciliation is that for a norm the messenger's visible membership in the reference group \emph{is} deep structure: her face is evidence about who holds the belief, in the way that a woman village leader \cite{beaman2009powerful} or a television character \cite{jensen2009power} is evidence. This also reconciles the split in the cultural-adaptation literature, where face-to-face adapted interventions show moderate benefit \cite{griner2006culturally} while adapted digital promotion often does not \cite{balci2022culturally}. Adaptation is not a uniform multiplier on health communication; it is an intervention on the messenger's social standing, and it should be expected to pay only where a belief's persistence is social.

\subsection{Cultural adaptation boosted effect size and durability}

Cultural adaptation roughly doubled the within-session shift ($+0.121$ against $+0.064$). Neither condition lost ground between the intervention session and three weeks. The culturally adaptive condition's advantage was therefore intact at follow-up and indeed at its widest. Past work has shown that single-session media effects on belief routinely decay within days \cite{ecker2022psychological}; however, ours survived.

\subsection{Designing culturally adaptive systems at scale}

Recent progress in the video generation space makes a messenger cheap enough to produce for every community that needs one, but our results suggest concentrating that effort rather than spreading it. Adaptation should be reserved for misinformation enforced by community expectation; misinformation that are simple knowledge gaps travel adequately without the need for a culturally adaptive AI video. Evaluation should not rely on perceived tailoring, since the
condition that changed belief more was not rated as more tailored, and formative testing that selects a variant on how well recipients say it speaks to them will select the wrong variant.

Two constraints qualify the scalability argument. Generators render non-Western subjects with uneven cultural faithfulness \cite{qadri2023ai, rani2026culturescoreevaluatingculturalfaithfulness}, making community validation a prerequisite rather than a refinement. And delivery here was necessarily mediated: $17\%$ of our sample cannot operate a phone unaided and a further $26\%$ need help, so self-serve distribution would exclude two-fifths of the population the intervention is for. The pipeline removes the cost of producing a culturally adaptive AI video; it does not remove the cost of getting it in front of the target audience.

\section{Limitations and Future Work}
\label{sec:limitations}

\textbf{Behavioral Changes}. While the claims we chose have direct behavioral counterparts such as antenatal registration and immunization uptake, for which administrative records exist, our efficacy assessments have relied on self-reported data. Other studies, such as the immunization experiments reviewed in Section~\ref{sec:delivery}, have observed vaccination records rather than stated belief \cite{banerjee2010improving, banerjee2025selecting}. Following belief change through to those records is the study that would establish whether this intervention is worth deploying.

\textbf{Group Viewing.} Participants watched the video in a group setting, typically alongside other women at the same literacy center, even though belief ratings and interviews were administered individually and privately (Section~\ref{sec:experimental_protocol}). A participant's stated 
belief could therefore have been shaped by visible reactions from other women in the room, rather than by the video alone. Several participants' own accounts point in this direction: our qualitative analysis found that women described hesitating to challenge a belief in front of others out of concern for how 
they would be seen (Section~\ref{sec:qual}, \emph{Social Hesitancy}), and one interview question asked directly how the room seemed to receive the video. We cannot separate the effect 
of the video itself from the effect of watching it alongside peers.

\section{Conclusion}

This study shows that a culturally adaptive AI video durably reduces belief in women's health misinformation: across 434 low-literacy women in Delhi, a culturally adaptive condition roughly 
doubled the immediate shift in belief over an identical script delivered by a culturally neutral video, and because neither condition decayed, its advantage was still intact three weeks later ($d=0.43$). The effect was concentrated, not uniform: it was largest on claims sustained by community norms rather than by gaps in medical knowledge, and it was largest, too, for women who 
described the \emph{grounds} on which they accepted a claim shifting, rather than simply receiving a new fact. And it was not something participants reported noticing: culturally adaptive condition participants did not describe feeling more personally addressed than neutral condition participants. What tracked belief change instead was whether participants took up the presenter as a named individual: women in the culturally adaptive condition referred to her by name or as a family-like figure, while no participant in the neutral condition ever used the name her video had given her, even though both videos introduced their presenter by name. As generated video makes a culturally adaptive AI presenter scalable to produce for any community, our 
findings point to a promising application for reaching traditionally underserved populations.

\section{Data Availability}
All data, code, and materials from this study are available on Hugging Face (https://huggingface.co/datasets/ankurani/healthvoices).

\section{LLM Usage Disclosure}
We used LLMs for minor writing assistance, including grammar correction and language
polishing. The core research ideas, methodology, experimental design, implementation, analysis, and conclusions were developed and carried out by the authors.

\begin{acks}
The authors would like to thank the Media Lab Consortium, the MIT Tata Center Technology and Design Fellowship, and the Google Cloud Credits award for supporting this work. The authors also thank Dr. Praveer Sinha (CEO \& MD, Tata Power), Dr. Ganesh Das (Chief Innovation \& Collaboration, Tata Power), and his team (Bharat Kumar Chhabra, Saptarshi Kar, Vinay Kumar, Pallavi Sagar, and Rupal), and our NGO partner representatives (Ritu Sharma, Anuradha, Jyoti Sharma, and Chanchal Aggarwal) for their help in conducting the field study responsibly. Finally, the authors thank Dr. Sana Parween for early feedback on the health questionnaire design, and Chanakya Ekbote, Twishmay Shankar, Shrihari Viswanath, and Sthit Parida for the discussions and feedback on the project.
\end{acks}

\bibliographystyle{ACM-Reference-Format}
\bibliography{_bib}


\newpage

\appendix

\section{Misinformation Spread and Impact}

\begin{figure*} [!htbp]
    \centering
    \includegraphics[width=\textwidth]{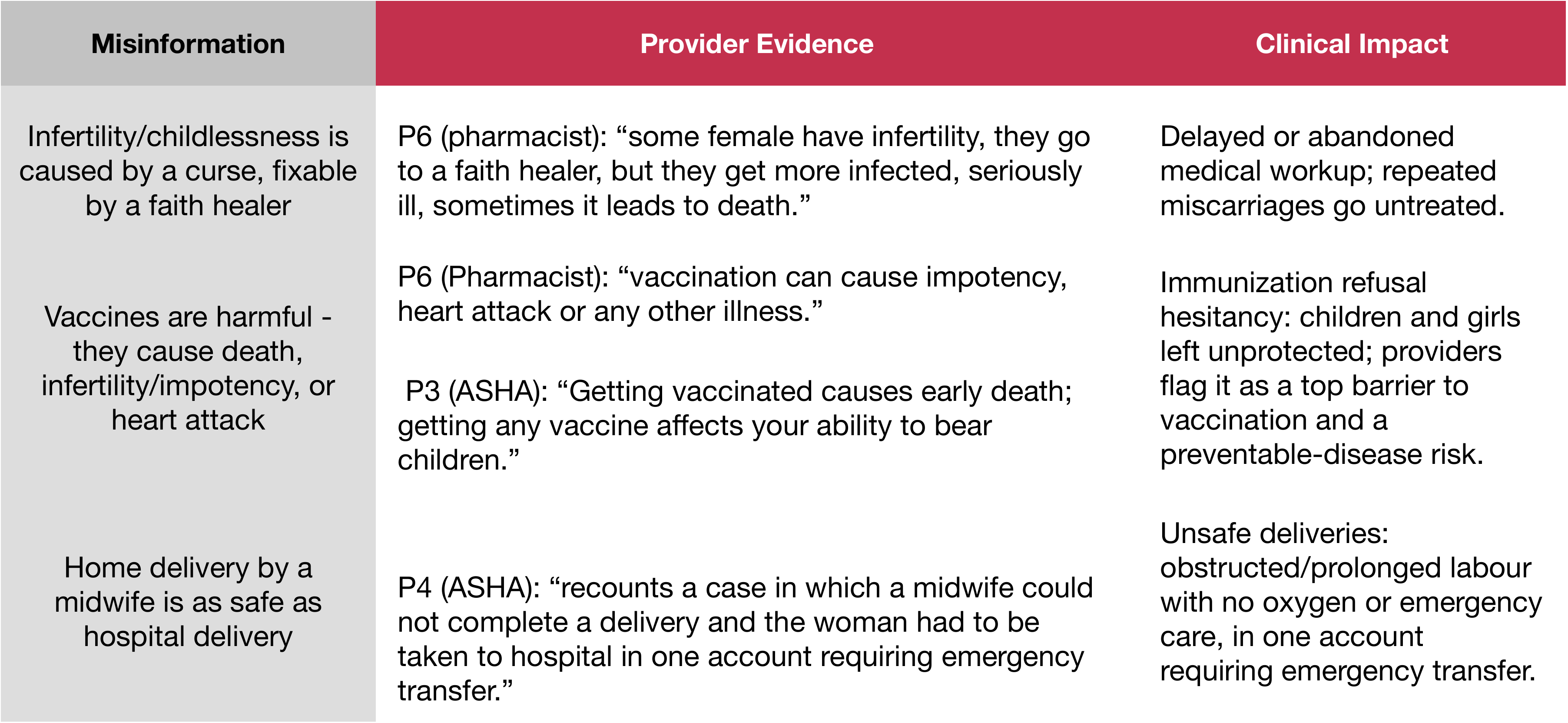}
    \caption{\textbf{Misinformation reported by healthcare providers.} Three examples from the field study, each with supporting provider evidence and the resulting clinical impact.}
    \label{fig:misinf-imp}
    \Description{A three-column table with headers Misinformation, Provider Evidence, and Clinical Impact, covering three beliefs: infertility is caused by a curse and fixable by a faith healer; vaccines cause death, infertility, or heart attack; and home delivery by a midwife is as safe as hospital delivery. Each row pairs the belief with quotes from pharmacists or ASHA workers and its clinical consequence, such as untreated miscarriages, vaccine refusal, and unsafe deliveries needing emergency transfer.}
\end{figure*}

We provide a list of misinformation, provider-reported clinical impact, and quotes from provider in Figure \ref{fig:misinf-imp}.

\section{Participant Demography}
\label{app:demographics}

\begin{figure*}[t]
  \centering
  \includegraphics[width=\textwidth]{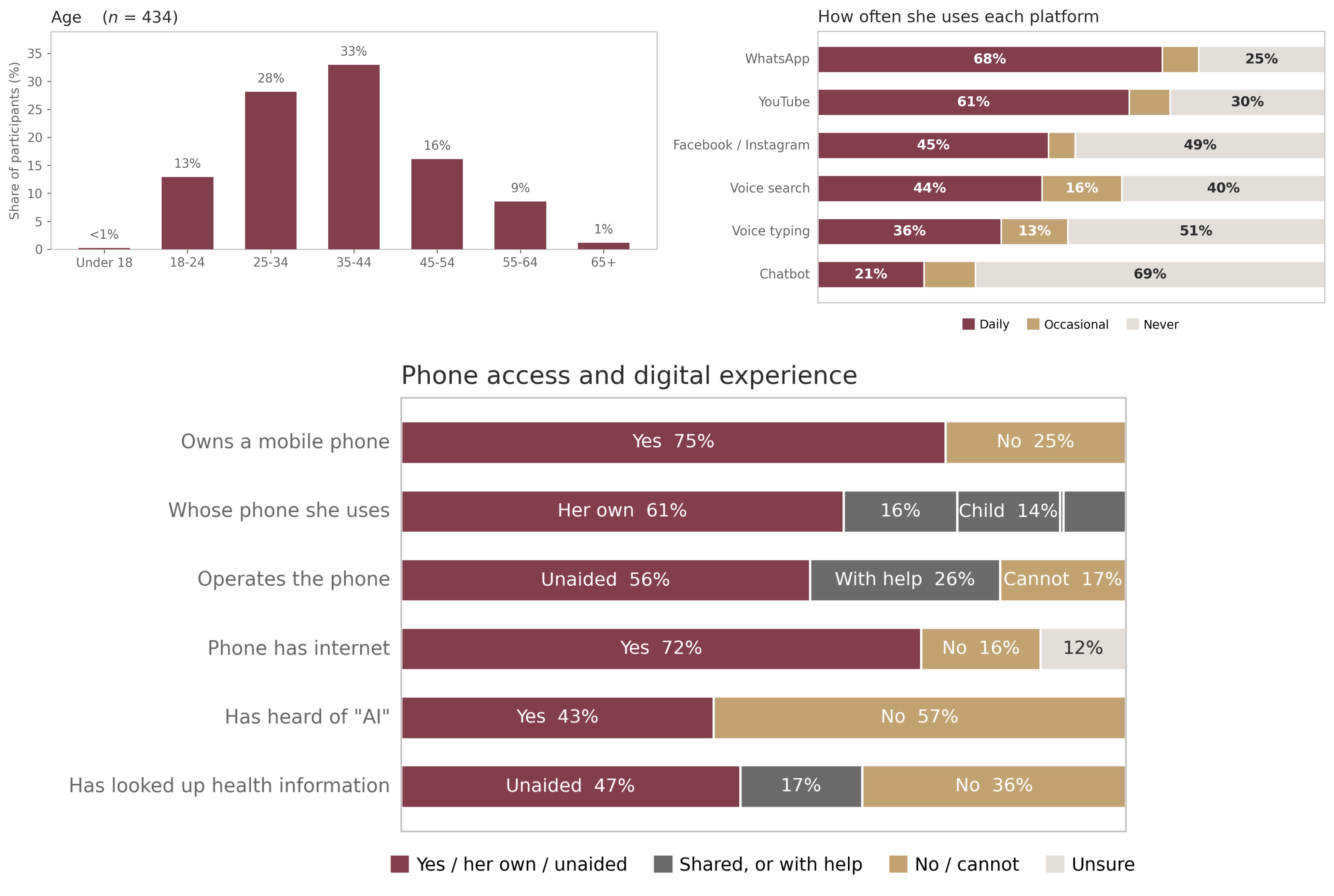}
  \caption{Demography details. Age distribution of the analyzed sample, pooled across the three conditions. Phone access, autonomy of use, and prior digital experience. ``Unaided'' denotes doing the task without help; ``with help'' with another person's assistance. Frequency of use of each platform, ordered by daily use. The survey's seven response bands are collapsed into daily use, occasional use (weekly or monthly), and never. Use falls away sharply for anything requiring composition or open-ended interaction.}
  \Description{Three charts describing participant demographics and digital experience. Top left, a bar chart of age shows most participants between 25 and 44, with few under 18 or over 65. Top right, stacked bars show how often participants use each platform, ordered by daily use: most use WhatsApp and YouTube daily, followed by Facebook or Instagram, voice search, and voice typing, while chatbots are rarely used. Bottom, stacked bars on phone access show that most participants own a phone, usually use their own, and have internet access, though some share a phone or need help operating it. Fewer than half have heard of AI, and about a third have never looked up health information.}
  \label{fig:demography}
\end{figure*}

Figures~\ref{fig:demography} report the demographic profile of the analyzed sample ($N = 434$), pooled across the three conditions. Percentages are of the women answering each item; item-level non-response was negligible, between zero and five women per item. Beyond the summary in Section~\ref{sec:participants}, three features are worth noting. Ownership overstates access: 61\% of the sample use a phone that is their own, the remainder using a husband's (16\%), a child's (14\%), or another household member's, so private viewing cannot be assumed. Connectivity is uncertain rather than absent --- 72\% report the phone has internet, while 12\% do not know whether it does. And health information seeking is already partly established, with 47\% having looked up health information on a phone unaided and a further 17\% having had someone do it for them, leaving 36\% who never have. The intervention therefore enters a population that is neither wholly outside digital health information nor reliably able to reach it alone.

\section{Field Interview Questionnaire} \label{app:interview}

\begin{figure*}[t]
  \centering
  \includegraphics[width=0.9\textwidth]{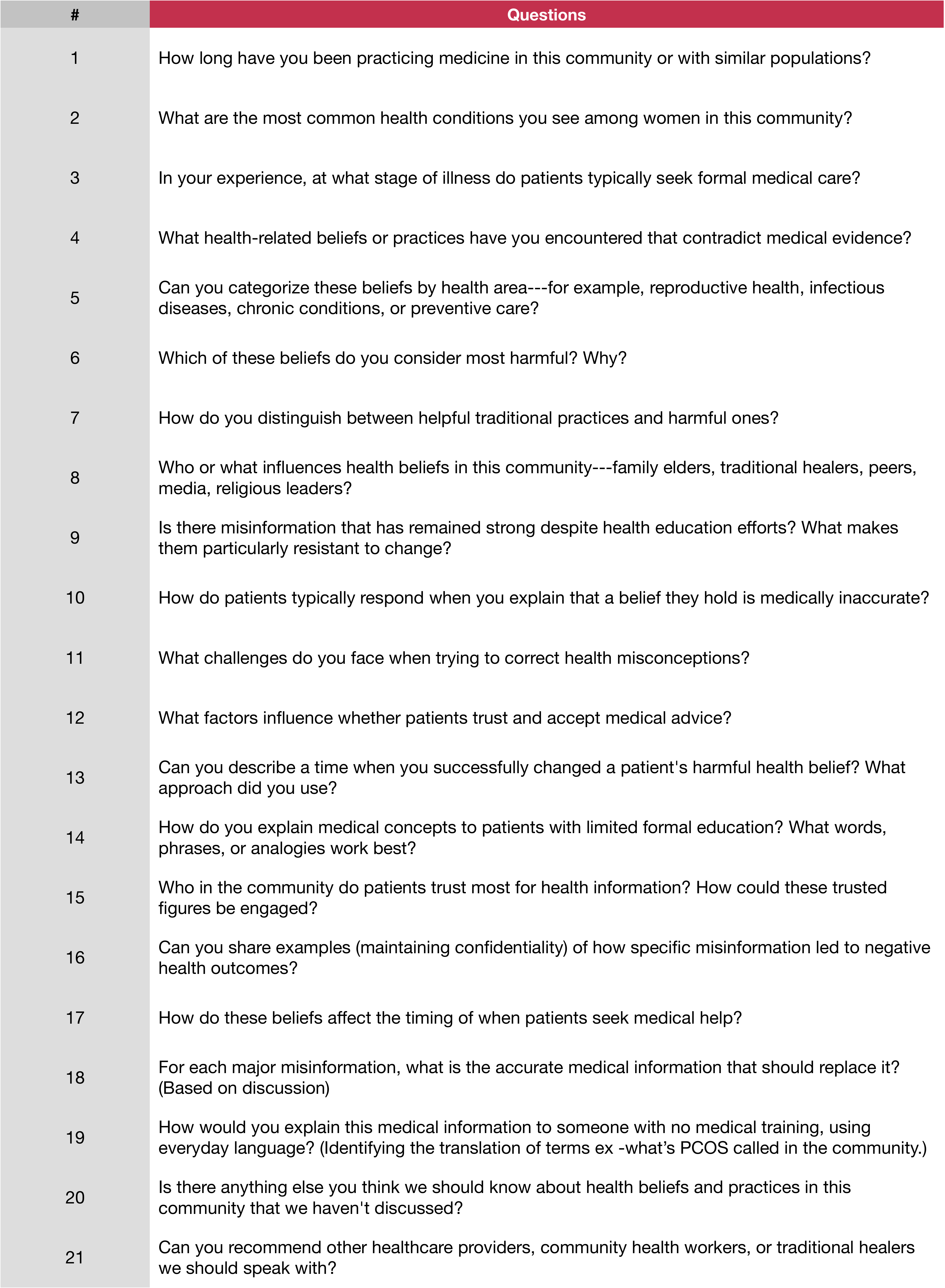}
  \caption{Field study Questions for curating a list of misinformation}
  \Description{A numbered table of 21 interview questions asked of healthcare providers in the field study. The questions cover the provider's experience in the community, common health conditions among women, and when patients seek care. They also ask which health beliefs contradict medical evidence, which are most harmful, and who shapes them. Later questions cover how patients respond to correction, what builds trust, how to explain medical concepts in everyday language, examples of misinformation leading to harm, and recommendations for other people to interview.}
  \label{fig:field-study-questions}
\end{figure*}

A field study was conducted to curate a list of misinformation that exists in the community, and the interview questions are presented in Figure \ref{fig:field-study-questions}.

\section{Prompt used for Generation}
\label{app:prompts}

\begin{figure*}
  \centering
  \includegraphics[width=\textwidth]{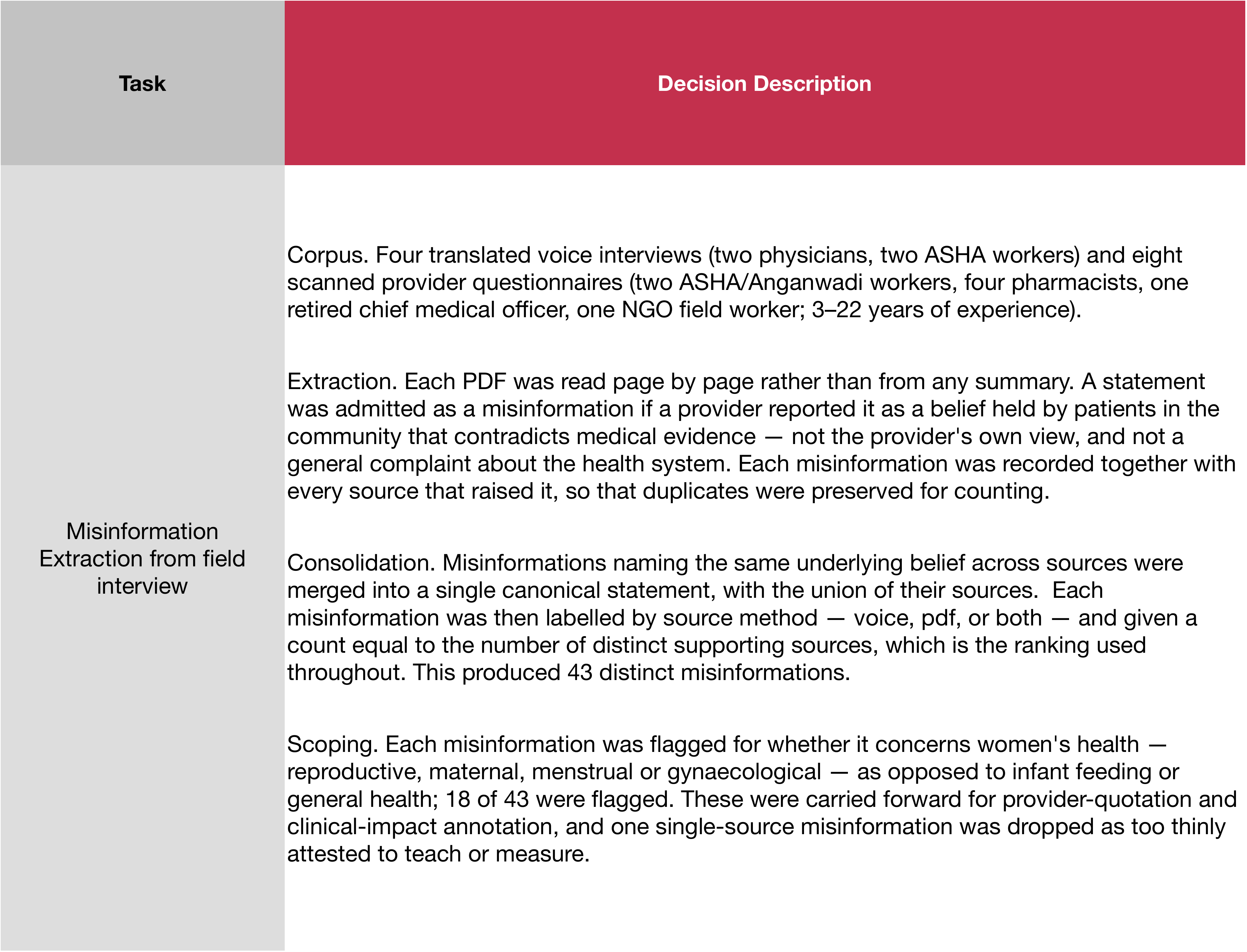}
  \caption{Decisions taken while extracting misinformation from interviews of healthcare providers.}
  \label{fig:decision}
  \Description{A two-column table describing the decisions made when extracting misinformation from healthcare provider interviews. It has four steps. Corpus: four translated voice interviews and eight scanned questionnaires from physicians, ASHA and Anganwadi workers, pharmacists, a retired chief medical officer, and an NGO field worker. Extraction: a statement was kept only if a provider reported it as a patient belief that contradicts medical evidence. Consolidation: duplicate beliefs across sources were merged, yielding 43 distinct misinformation statements ranked by number of sources. Scoping: 18 statements concerning women's health were retained, and one thinly supported statement was dropped.}
\end{figure*}

\begin{figure*}
  \centering
  \includegraphics[width=\textwidth]{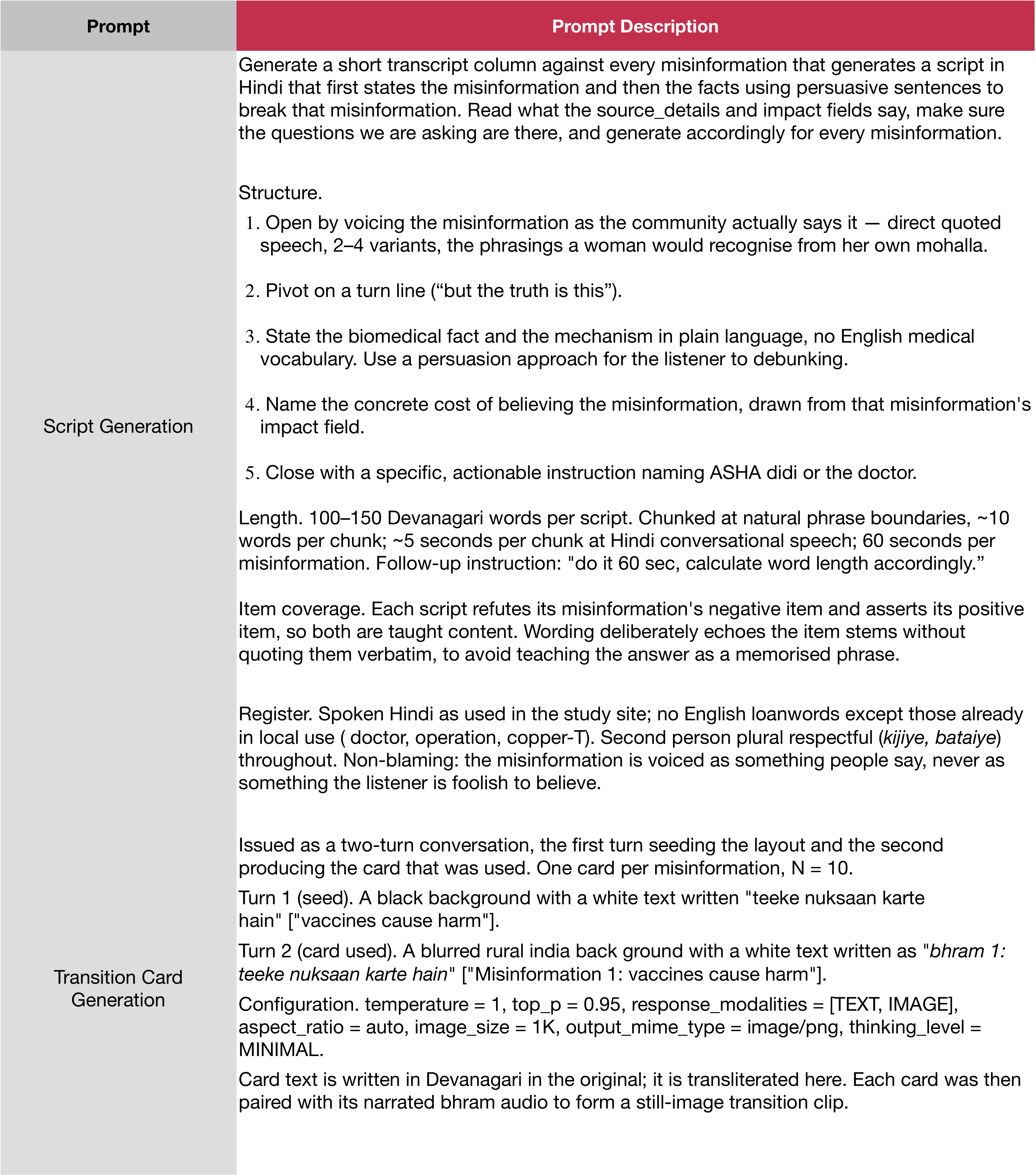}
  \caption{Prompts for script and transition card generation.}

  \Description{A two-row table of prompts used to generate the video content. The first row, script Generation, describes a prompt that writes a Hindi script for each misinformation claim. Each script opens with the claim as the community phrases it, pivots to the truth, explains the medical fact in plain language, states the cost of believing the claim, and closes by directing the listener to an ASHA worker or doctor. It also specifies script length, item coverage, and a respectful, non-blaming spoken register. The second row, Transition Card Generation, describes a two-turn image prompt that produces one title card per claim, showing the claim in Devanagari text over a blurred rural background, along with the model configuration used.}

  \label{fig:prompt-dialogue}
  
\end{figure*}

\begin{figure*}
  \centering
  \includegraphics[width=\textwidth]{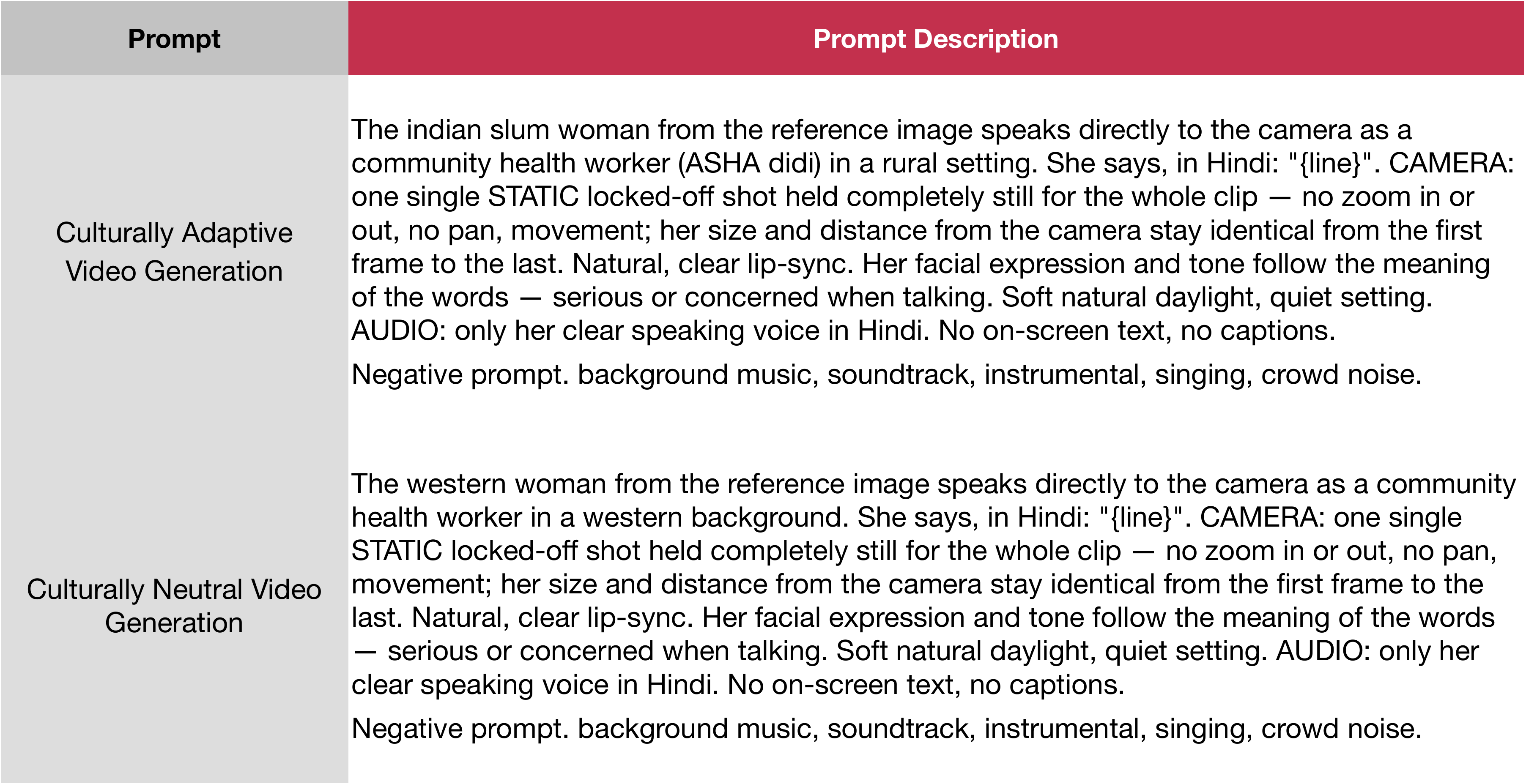}
  \caption{Prompts for image-conditioned video generation}

  \Description{A two-row table of the text prompts used to generate the presenter videos from reference images. In the culturally adaptive prompt, an Indian woman speaks to the camera in Hindi as an ASHA community health worker in a rural setting. In the culturally neutral prompt, a Western woman speaks the same Hindi line as a community health worker in a Western setting. Both prompts specify a single static shot, natural lip-sync, expressions matching the words, soft daylight, and only the speaker's voice. Both also share a negative prompt excluding background music, singing, and crowd noise.}
  
  \label{fig:prompt-video}
\end{figure*}

In this section, we describe the decisions used for extracting the list of misinformation (Figure \ref{fig:decision}), prompts for script and transition card generation (Figure \ref{fig:prompt-dialogue}), and for both conditions of video generation (Figure \ref{fig:prompt-video}).

\section{Generated Scripts and Practitioner-Revised Script} 
\label{app:script_dialogue}

We verified the generated scripts from health care providers and present three English-translated pairs in Table \ref{tab:script-revision}.

\begin{table*}[t]
  \centering
  \footnotesize
  \begin{tabularx}{\textwidth}{@{}p{0.15\textwidth} Y Y@{}}
    \toprule
    \textbf{Misinformation} & \textbf{Generated script (short transcript)} & \textbf{Practitioner-revised script (filmed)} \\
    \midrule

    Infertility/childlessness is caused by a curse or God's anger and can be fixed by prayers
    or a \emph{baba} (faith healer), instead of medical care &
      When a woman does not conceive, the talk starts in the \emph{mohalla} ---
      ``She is under a curse.'' ``God above is angry.'' ``Get \emph{jhaad-phook} done by a
      \emph{baba}.'' But the truth is this --- childlessness has a physical cause. Sometimes a
      hormonal imbalance, sometimes a blockage in the tubes, sometimes a problem in the man.
      Every cause can be tested for, and most can be treated. But this testing can only happen
      at a doctor's. Doctors tell us --- in the rounds of the \emph{babas} women fall even more
      ill, infection spreads, and sometimes lives are lost. The longer the \emph{jhaad-phook}
      goes on, the longer the real treatment is held up. A curse is not an illness. Do not
      delay --- husband and wife, both of you, get tested by a doctor. &
      ``When a couple does not have a child, people begin to say all sorts of things --- `It is
      a curse', `God is angry', or `Get \emph{jhaad-phook} done by a \emph{baba}.' ''\par
      But there can be a reason connected to the body behind not having a child. Sometimes a
      hormonal problem, sometimes a blockage in the tubes, and sometimes the difficulty can be
      from the husband's side as well.\par
      The good thing is that these causes can be tested for, and in many cases treatment is
      possible too. Spending time on \emph{jhaad-phook} can delay the start of the right
      treatment.\par
      So do not delay. Husband and wife, both together, meet a doctor and get the necessary
      tests done. Only when the right cause is known can the right treatment be given. \\
    \addlinespace

    Vaccines are harmful --- they cause death, infertility/impotency, or heart attack &
      You too must have heard --- ``Do not get the girl vaccinated; later on she will not be
      able to have children.'' Or --- ``The vaccine takes lives.'' But the truth is this --- a
      vaccine teaches the body to fight disease. It has no effect on the ability to become a
      mother. Across the world, crores of vaccinated girls are healthy mothers today. And the
      tetanus vaccine given to a pregnant woman saves the lives of both mother and newborn.
      When a vaccine is missed out of fear of a rumor, the child is left exposed to measles,
      polio and tetanus --- which were very easy to be protected from. The price of one missed
      vaccine may have to be paid for a whole lifetime. Trust your \emph{ASHA didi} and the
      doctor, not the talk of the \emph{mohalla}. &
      ``You too must surely have heard things like this in your \emph{mohalla} at some time or
      other\ldots''\par
      ``Do not get the girl vaccinated; later on she will not have children\ldots'' ``Getting
      the vaccine is a danger to life\ldots''\par
      But are these things true? Absolutely not!\par
      Getting vaccinated has no effect on a girl's ability to become a mother later on. On the
      contrary, vaccines give our body the strength to fight many serious diseases.\par
      And when it comes to pregnancy, the necessary vaccines given on time help protect both
      the mother and the coming child.\par
      Many times we trust the talk around us, or rumors, and avoid getting a necessary
      vaccine. But doing this can be harmful, because it can mean losing the chance of
      protection from serious diseases like measles, polio and tetanus.\par
      So the next time someone says to you --- ``Do not get her vaccinated, she will not have
      children later\ldots'' --- then before trusting such talk, be sure to ask your
      \emph{ASHA didi} or the doctor.\par
      Trust correct information, not rumors. Get the necessary vaccines on time, and keep
      yourself and your family safe. \\
    \addlinespace

    Home delivery by a \emph{dai} (traditional birth attendant) is safe/acceptable &
      The elders of the house say --- ``The \emph{dai} delivered all the babies in our time; at
      home is just as safe as the hospital.'' But in a delivery anything can happen suddenly ---
      the baby can get stuck, there can be heavy bleeding, or the baby's breathing can stop. At
      such a time, every single minute is precious. In a hospital, there is an arrangement for
      oxygen, blood, and an operation. At home there is no such arrangement. Our \emph{ASHA didi}
      tells us --- one woman's labor pains went on for hours, the \emph{dai} could do nothing,
      and her life was saved only when she was rushed to the hospital. Many \emph{dais} have
      neither a degree nor equipment. Have the delivery in a hospital. Decide in advance on a
      vehicle, money, and who will go with you. &
      ``The elders of the household often say --- `In our time the \emph{dai} used to do the
      delivery right at home, and nothing happened to us.' ''\par
      But today there is one thing we must understand --- there is no way of knowing in advance
      when some trouble may suddenly arise during a delivery.\par
      Sometimes the baby can get stuck, sometimes there can be very heavy bleeding, or sometimes
      the baby can have difficulty breathing. At such a time, every single minute is very
      precious.\par
      In a hospital, facilities such as oxygen, an arrangement for blood, and an operation if
      needed are available to handle such a situation. At home, these facilities are hard to come
      by.\par
      So try to make sure the delivery happens in a hospital.\par
      And not only deciding on the hospital --- also think in advance about how you will get to
      the hospital, who will go with you, and how you will keep some money ready for an
      emergency.\par
      Remember, a little preparation done in advance can prevent a very big problem, and can
      help protect both the mother and the baby.\par
      So keep your preparations complete before the delivery, and give priority to a hospital
      for a safe birth. \\
    \bottomrule
  \end{tabularx}
  \caption{Three of the ten scripts, showing the model-generated 60-second script and the version after review by community health practitioners and a social-impact field team. The revised column is what was chunked and passed to the video generator. Scripts were written and filmed in Hindi; they are translated here, with community terms retained.}
  \Description{Table comparing generated scripts with practitioner-revised versions for three misinformation, in English translation.}
  \label{tab:script-revision}
\end{table*}

\section{Experimental Protocol Questions}

We also present a list of all questions used to ask participants across all phases in Table \ref{tab:item-schedule}

\begin{table*}
\centering\small\setlength{\tabcolsep}{5pt}
\begin{tabular}{@{}p{5.9cm}cccc@{}}
\toprule
 & \multicolumn{2}{c}{\textbf{Phase 1}} & & \\
\cmidrule(lr){2-3}
\textbf{Question} & pre & post & \textbf{Phase 2} & \textbf{Phase 3} \\
\midrule
Can childlessness have a physical cause that a faith healer could diagnose? & \checkmark & \checkmark & \checkmark & \checkmark \\
Is family planning (condoms, pills, IUCD, injections) safe for women's health? & \checkmark & \checkmark & \checkmark & \checkmark \\
Is using old cloth during periods less safe than a sanitary pad? & \checkmark & \checkmark & \checkmark & \checkmark \\
Does the fault in childlessness lie only with the woman? & \checkmark & \checkmark & \checkmark & \checkmark \\
Can pickle spoil if touched by a woman during her period? & \checkmark & \checkmark & \checkmark & \checkmark \\
Are some vaccinated girls later unable to have children? &  & \checkmark & \checkmark & \checkmark \\
Can repeated miscarriage be cured by exorcism? &  & \checkmark & \checkmark & \checkmark \\
Does telling the ASHA \emph{didi} or doctor in the first three months benefit mother and child? &  & \checkmark & \checkmark & \checkmark \\
Can the problem of childlessness lie with the man, and does it need investigating? &  & \checkmark & \checkmark & \checkmark \\
Can bleeding after menopause signal a serious illness such as cancer? &  & \checkmark & \checkmark & \checkmark \\
Is delivering at home with a midwife equivalent to delivering in hospital? &  &  & \checkmark & \checkmark \\
Does a pregnant woman drinking milk cause gas in the baby, so she should avoid it? &  &  & \checkmark & \checkmark \\
Is a sanitary pad safer than old cloth during periods? &  &  & \checkmark & \checkmark \\
Can pregnancy still occur after sterilization? &  &  & \checkmark & \checkmark \\
Is menstrual blood normal blood, not a sign of impurity? &  &  & \checkmark & \checkmark \\
Do vaccines in pregnancy protect both mother and unborn child? &  &  &  & \checkmark \\
Do milk, greens and nutritious food keep mother and baby healthy? &  &  &  & \checkmark \\
Is it the case that childbirth cannot be safely arranged at home? &  &  &  & \checkmark \\
Is bleeding after menopause due to heat or weakness, and does it resolve on its own? &  &  &  & \checkmark \\
Does telling about a pregnancy in the first three months risk miscarriage? &  &  &  & \checkmark \\
\bottomrule
\end{tabular}
\caption{The twenty belief items and the phases at which each was administered. Each claim is measured by two differently worded items administered at different phases, so no phase repeats a claim's wording, and each phase includes questions the participant has not seen before. Hindi originals are in the released materials.}

\Description{A table of twenty belief items about women's health, marking the phases at which each was asked: Phase 1 before and after the video, Phase 2, and Phase 3. Five items are asked at Phase 1 before the video, and five new items are added at each later point; each phase repeats all earlier items, and Phase 3 includes all twenty.}
\label{tab:item-schedule}
\end{table*}

\subsection{Qualitative questions}
\label{sec:qual-questions}

\begin{figure*}
  \centering
  \includegraphics[width=\textwidth]{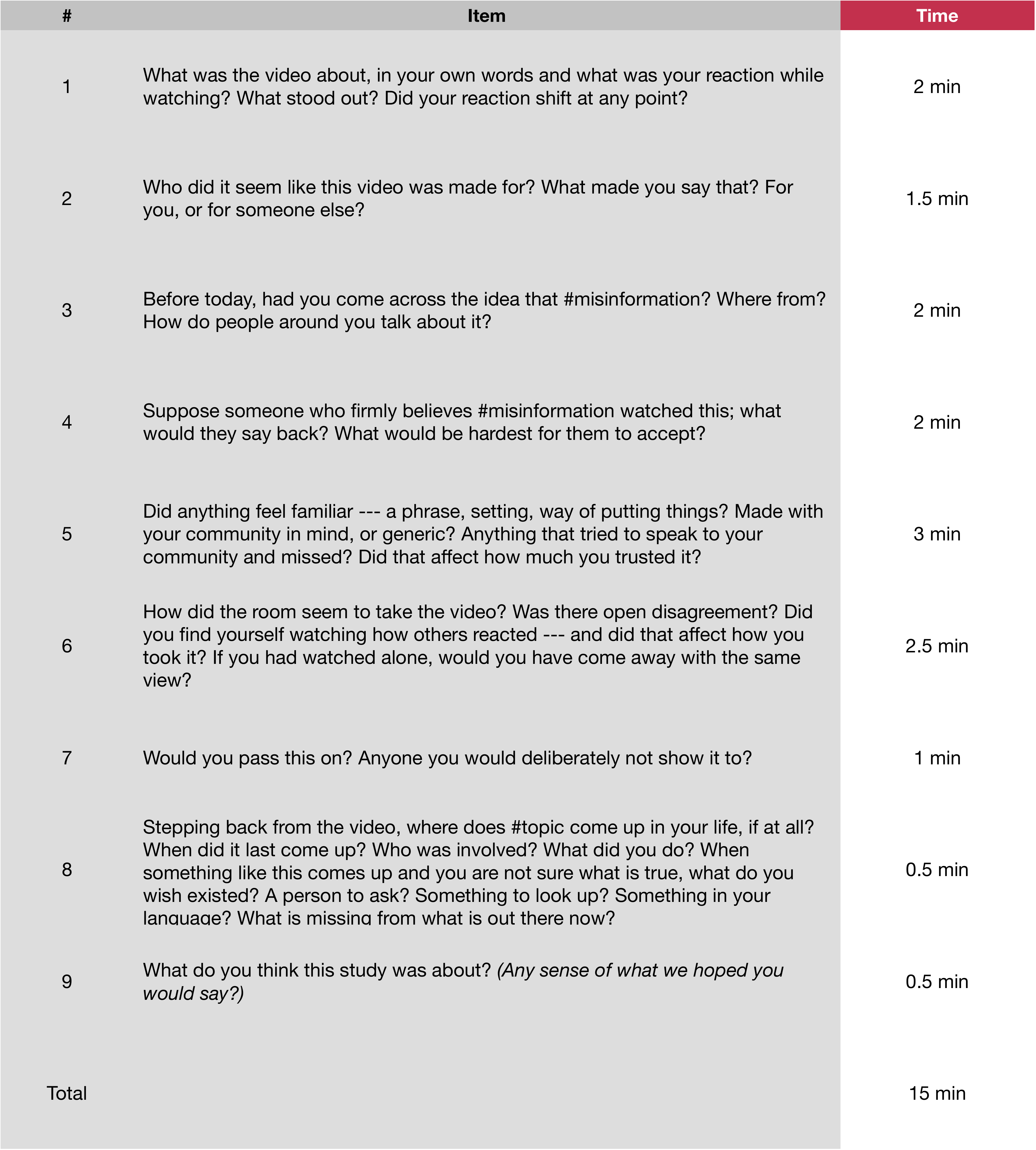}
  \caption{Post-exposure interview schedule (15 minutes per participant). Administered in the same session, individually. Identical in both arms.}
  \Description{A table listing nine post-exposure interview questions with the time allotted to each, totaling 15 minutes. The questions ask participants to describe the video and their reaction, who they think it was made for, whether they had encountered the misinformation before, and how a firm believer would respond. They also ask whether the video felt familiar or made for their community, how others in the room reacted, whether they would share it, where the topic arises in their own life and what resources they wish existed, and what they think the study was about.}
  \label{fig:interview-schedule}
  
\end{figure*}

We present the list of qualitative questions in Figure \ref{fig:interview-schedule}.

\end{document}